# Laser-Induced Optical Reconfiguration in Ge–Sb–Te Films with Composition-Dependent Response

M. Zhezhu[1, †], A. Vasil'ev[1], M. Sukiasyan[2], A. Kutuzyan[2], N. Anosov[3], M. Yaprintsev[3], D. A. Ghazaryan[4], H. Gharagulyan[1, 2 †]

[1] *Liquid Crystalline Nanosystems Research Group, A.B. Nalbandyan Institute of Chemical Physics NAS RA, Yerevan 0014, Armenia*

[2] *Institute of Physics, Yerevan State University, Yerevan 0025, Armenia*

[3] *Belgorod State University, Belgorod 308000, Russia*

[4] *Laboratory of Advanced Functional Materials, Yerevan State University, Yerevan 0025, Armenia*

[†] *The correspondence should be addressed to:*
*marina.zhezhu@ichph.sci.am and herminegharagulyan@ysu.am*

**Abstract**

Phase-change Ge–Sb–Te (GST) materials exhibit pronounced optical contrast and tunability driven by structural transformations, enabling a diverse range of photonic and optoelectronic applications. GST materials undergo reversible amorphous–crystalline phase transitions during which the refractive index and extinction coefficient increase significantly in the crystalline phase across a broad spectral range. However, a systematic correlation between local composition, crystalline microstructure, and broadband optical response within as-deposited crystalline GST films has not been established, particularly for films spanning various compositions within a single growth process and in the absence of amorphous-crystalline transitions. Here, we report a systematic study of composition-resolved opto-structural property relationships in as-deposited crystalline GST films spanning $Ge_3Sb_2Te_6$, $Ge_2Sb_2Te_5$, and $GeSb_2Te_4$ within a single CVD process, avoiding intermediate phase transitions. This enables direct correlations between composition, microstructure, morphology, and broadband optical response across 400-1700 nm. Specifically, the compositional gradient results in a shift of the absorption minimum from ~815.9 nm to ~889.8 nm, accompanied by a 5.9-fold change in the intensity, reflecting the strong dependence of optical behavior on composition and microstructure. We further show that femtosecond laser irradiation enables spatially selective tuning of the optical response. Local laser irradiation induces fluence-dependent changes in the optical response across the Vis-NIR range. $GeSb_2Te_4$ exhibits the strongest modulation in the NIR range (absorption ratio is 3.2 at 191 J/cm$^2$), $Ge_2Sb_2Te_5$ shows the highest optical contrast in the Vis range (absorption ratio is 8.1 at 179 J/cm$^2$), whereas $Ge_3Sb_2Te_6$ exhibits a decrease in absorbance across both spectral ranges. These findings establish crystalline GST films as a platform for broadband-tunable, spatially programmable photonic elements, controllable through both compositional design and localized laser processing.

**Keywords:** phase change materials, chemical vapor deposition, broadband optical tuning, laser-induced structural transformation, spatial programmable optical materials, reconfigurable photonics

## Introduction

Phase-change materials (PCMs) based on Ge–Sb–Te (GST) alloys have been extensively investigated due to their ability to undergo reversible transitions between amorphous and crystalline states, which are accompanied by significant changes in electrical and optical properties [1, 2]. Their unique characteristics make GST compounds promising for applications in optical data storage [3], reconfigurable and non-volatile photonic systems [4], next-generation microelectronics [5, 6], metamaterials and metasurfaces [7], IR sensing and modulation devices [8], as well as reversible near-IR windows [9]. Most optical studies of GST materials have traditionally focused on the contrast between amorphous and crystalline states, where optical variations arise from differences in electronic structure, bonding configuration, and free-carrier density [10-14].

GST materials exhibit remarkable optical properties across both the visible and IR spectral ranges. Particularly, studies report that they exhibit the highest extinction coefficient in 400-1500 nm spectral range [5, 15]. In the visible region, GST-based multilayered structure enables functionalities such as camouflage and tunable reflectivity [16]. In the IR range, these materials demonstrate efficient absorption and photodetection capabilities, including in their crystalline phases [17-19]. The sensitivity of optical reflectivity to structural order has been demonstrated in $Ge_2Sb_2Te_5$ films, where variations in crystalline phase and degree of order result in measurable changes in reflectance across the visible spectral range [20, 21]. Similarly, Ge-rich GST compositions such as $Ge_{14}(Sb_2Te_3)_{86}$ exhibit pronounced reflectance variations in the NIR region, enabling multi-level optical modulation and wavelength-selective behavior [22]. Moreover, the optical constants of $Ge_2Sb_2Te_5$ have been demonstrated to be finely tunable *via* laser irradiation methods [15]. Previous work demonstrated that CVD-grown crystalline GST films with lateral compositional gradients act as passive IR filters, with $Ge_2Sb_2Te_5$ region showing strong absorption and $Ge_3Sb_2Te_6$ or $GeSb_2Te_4$ regions showing weaker absorption [23]. This finding highlights the strong coupling between composition, bonding configuration, and optical response in crystalline GST alloys, and enables gradient optical filtering without complex lithographic processing.

However, studies on the optical properties of as-grown crystalline GST pseudo-binary compositions remain relatively underexplored. Most reported studies on the Vis-IR optical properties of GST materials have relied on the traditional approach of depositing amorphous films followed by thermally/laser-induced crystallization [10, 11, 14, 24, 25]. In such cases, the crystallization pathway itself can play a decisive role in determining the final structural state and, consequently, the optical response [5, 15, 26-29]. Given the intrinsic multifunctionality of GST films, it is therefore important to understand how crystalline GST grown materials behave across a broad spectral window and how their optical response can be engineered through composition and morphology.

Beyond intrinsic control through material design, post-growth modification provides an additional route for tailoring the optical properties of crystalline films. In this context, the active and spatially selective tuning of optical properties *via* ultrafast laser irradiation has not yet been systematically explored in such systems. Consequently, the interplay between laser-induced structural transformations, compositional gradients, and broadband visible–IR optical response remains insufficiently understood. Addressing this gap is essential for the development of GST-based reconfigurable photonic platforms, adaptive infrared optical components, and spatially programmable functional devices.

In this work, we systematically investigate opto-structural property relationships in as-deposited crystalline Ge-Sb-Te films with an in-plane compositional gradient. Crystalline $Ge_3Sb_2Te_6$, $Ge_2Sb_2Te_5$, and $GeSb_2Te_4$ compositions were synthesized within a single CVD process. This approach enables a direct correlation between composition, microstructure, and intrinsic optical response without intermediate phase transition. We first establish the baseline relationships between composition, morphology, and broadband Vis to NIR (400–1700 nm) optical properties using Raman spectroscopy, scanning electron microscopy (SEM), and optical spectroscopy. Subsequently, we examine the evolution of the optical response under femtosecond laser irradiation, focusing on spatially localized post-growth modifications. We demonstrate that composition-dependent structural and morphological features strongly govern the initial optical behavior, while laser-induced modifications introduce additional opportunity for spatially selective tuning of the optical response. Overall, this work establishes a platform for controlled, spatially programmable tuning of broadband optical properties in GeTe-$Sb_2Te_3$ pseudo-binary materials, enabling their application in reconfigurable photonic systems and adaptive optical devices.

**1. Experimental Section**

Gradient crystalline Ge–Sb–Te films were deposited *via* CVD simultaneously in a single process by varying the source-to-substrate distance without changing the precursor. For this, a source material was first synthesized from high-purity elemental Ge (99.99%), Sb (99.999%), and Te (99.999%) by direct melting in an evacuated and sealed quartz ampoule. The elements were loaded into the ampoule in the stoichiometric ratio Ge:Sb:Te = 2:2:5. The ampoule was gradually heated to 1000 °C, held at this temperature for 6 h to ensure complete homogenization of the melt, and then slowly cooled to room temperature to obtain the $Ge_2Sb_2Te_5$ source material. A fragment of the synthesized bulk alloy crystal $Ge_2Sb_2Te_5$ with a mass of approximately 100 mg was placed in a quartz tube reactor (50 mm inner diameter). Glass cover slips (18 mm × 18 mm, thickness ~0.15 mm) were used as substrates. The distance between the $Ge_2Sb_2Te_5$ source and the first glass substrate was set to 10.5 cm. Two additional substrates were placed sequentially behind the first, forming a linear arrangement of three substrates. During deposition, a constant vacuum pressure of 10 Pa was maintained using a rotary vacuum pump without introducing a carrier gas. The reaction tube was heated to 550°C over 30 minutes and held at this temperature for an additional 30 minutes. To terminate the deposition process, the quartz tube was quickly removed from the heating zone. The deposited GST films were labeled according to their position relative to the source: Film 1 corresponds to the farthest substrate, Film 2 to the middle substrate, and Film 3 to the closest substrate. The deposition parameters are described in detail in our previous work [23], and a schematic representation is shown in Figure S1.

The morphology and microstructure of the films were examined using SEM equipped with an energy-dispersive X-ray spectroscopy (EDXS) kit for elemental analysis. SEM images provided insight into the surface features, films' thicknesses, while EDXS confirmed the spatial variation of Ge, Sb, and Te content across the films. Raman scattering measurements were performed to analyze the structural properties of the films before and after laser irradiation. Raman spectra were collected from different compositional regions and laser-irradiated spots using a Raman spectrometer (LabRam HR Evolution, Horiba) equipped with a 633 nm excitation laser (0.85 mW), a 600 gr/mm grating, an acquisition time of 10 s, and 3 accumulations. The choice of a 633 nm excitation wavelength ensures a favorable balance between Raman signal intensity and suppression of fluorescence background, making it the most widely used in studies of phase-change materials [30].

The optical properties were investigated in the Vis-NIR spectral range using spectrophotometers by StellarNet, namely, Black-Comet-SR (200-1100 nm) and Dwarf-Star InGaAs-512 (900-1700 nm). To estimate the optical bandgap of the films, absorption spectra were collected in the 400–1700 nm range. To avoid possible transition zones near the film edges (where the gradient may be ill-defined or the film thickness varies rapidly), optical spectra were collected from the center region of the gradient film. The incident wavelength (λ) and corresponding photon energy ($E = h\nu$) were then used to determine the optical band gap using the Tauc relation.

To analyze the optical behavior of in-plane gradient as-deposited Ge-Sb-Te films, absorbance measurements were performed in the 400-1000 nm spectral range along two principal directions. The longitudinal direction, $L$-direction (i.e., source-to-substrate direction), was defined as the axis from the source toward the substrate (representing the deposition gradient), whereas the transverse direction, $T$-direction, was oriented perpendicular to the deposition axis. Measurements were taken along the central line of the films with a spatial step of 2 mm in both directions.

Absorption spectra were recorded after localized laser irradiation in the 400-1700 nm spectral range to enable evaluation of laser-induced changes in optical absorbance.

Films were irradiated using a Coherent MIRA 900 Ti-Sapphire Laser operating at a wavelength of 800 nm and exhibiting a Gaussian intensity distribution. Laser exposure was focused on targeted areas of the films to induce local modification of optical and structural properties. The pulse duration, repetition rate, laser power, and irradiation time were systematically varied. The films were divided into distinct regions, each of which was exposed to multiple femtosecond laser pulses. The pulse numbers, corresponding to both repetition rates and all irradiation times, as well as the laser powers and pulse energies, are provided in SI (Table S1). The lowest fluence, 27 J/ cm$^2$, was achieved using 100 fs pulses at a repetition rate of 76 MHz and a power of 137 mW, corresponding to a pulse energy of 1.8 nJ. The highest fluence was achieved under the same pulse duration and repetition rate with a power of 700 mW and a pulse energy of 9.2 nJ, corresponding to a fluence of 4180 J/cm$^2$. The $GeSb_2Te_4$ film was exposed to femtosecond laser irradiation with fluences of 27, 82, 123, and 191 J/cm$^2$. The $Ge_2Sb_2Te_5$ film was irradiated using a laser with a fluence of 96, 116, 179, and 227 J/cm$^2$. For the $Ge_3Sb_2Te_6$ film, fluences varied as 55, 95, 164, 273, 697, and 4180 J/cm$^2$. Additional irradiation experiments were performed using a fluence of 0.38 J/cm$^2$, but with a longer single pulse (0.5 ms) and a high-intensity laser source (764 W/cm$^2$, compared to 27-139 W/cm$^2$ for the ultrafast laser). For this purpose, the films were irradiated under vacuum using the laser flash module integrated into the TC-1200RH thermal constant measurement system (Advance Riko, Inc.). The reported fluence values correspond to the cumulative (average) energy density delivered to the irradiated area during the entire irradiation period. The cumulative fluence was calculated as $F_{av} = N \cdot F_{pulse}$, where $N = f \cdot t$ is the total number of laser pulses, $f$ is the laser repetition rate (Hz), and $t$ is the irradiation time (s), and $F_{pulse} = \frac{E_{pulse}}{A}$ is the single-pulse fluence (J/cm$^2$), with $E_{pulse} = \frac{P}{f}$ , $P$ and $A$ being laser power (W) and spot area (cm$^2$), respectively.

## 2. Results and discussion

### 2.1 Structural Analysis of As-Deposited Films

The surface morphology and XRD analysis of the as-deposited GST films are presented in Figure 1. SEM analysis revealed that CVD-grown films deposited on glass substrates exhibit characteristic surface features. The films are composed of plate-like structures. This morphology results in a textured surface with clearly visible lamellar grain morphology. The spatial variation in crystallite size along the series of substrates correlates with their position relative to the hot zone, in agreement with the expected temperature gradient along the reaction tube. The grain size of Film 1 was estimated to be 0.2 µm. Film 2 exhibited more pronounced plate-like crystallites with a thickness of 0.3 µm and a length of 1.4 µm. Film 3 showed the largest plates, with a thickness of 0.4 µm and a length of 3.6 µm. Film thicknesses ranged from 3.8 $\mu$m for Film 1, 1.0 $\mu$m for Film 2, and 0.3 $\mu$m for Film 3. The corresponding cross-sectional SEM images are presented in Figure S2(a).

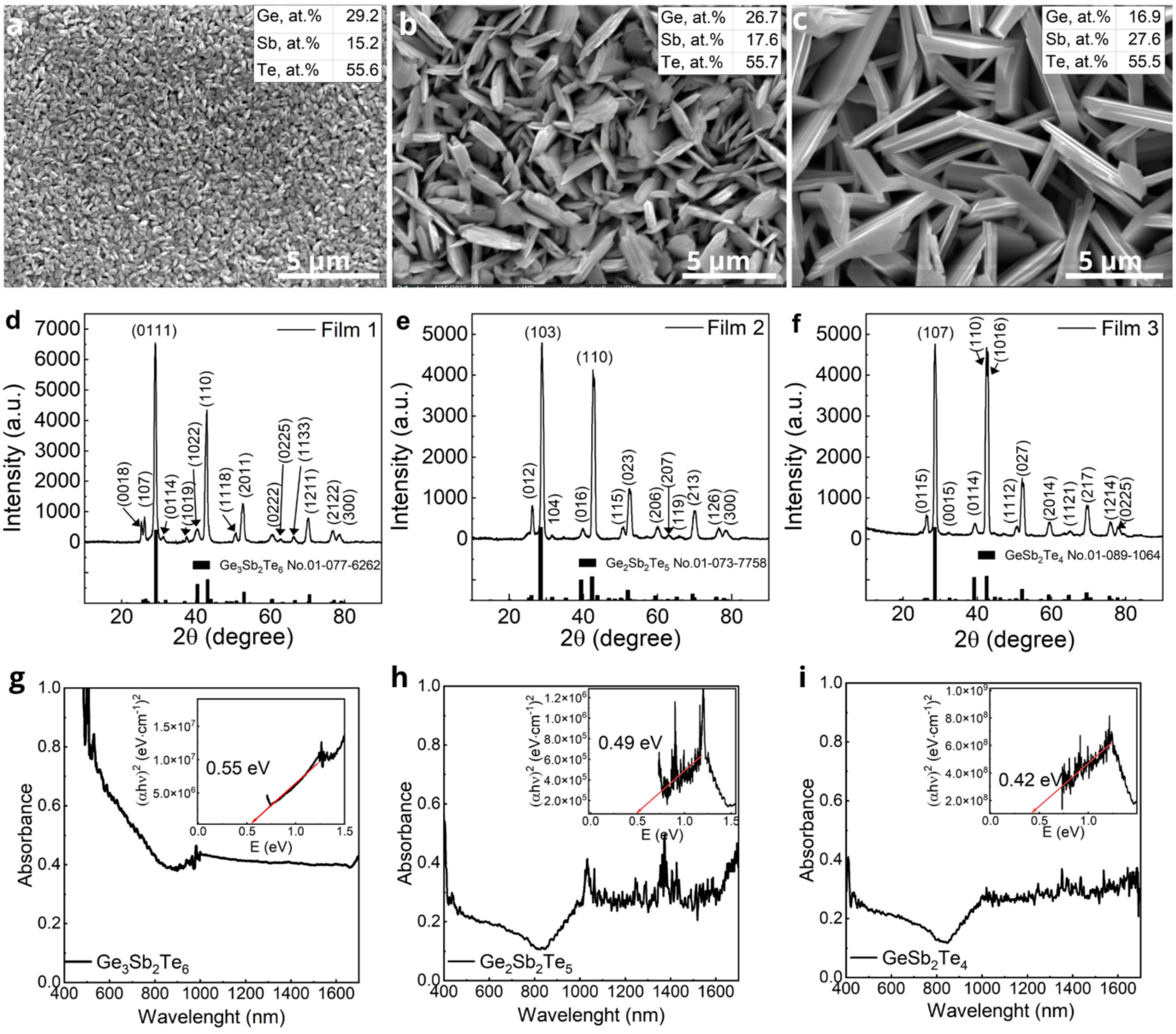

Figure 1. SEM images of CVD-grown Ge–Sb–Te films on glass substrates showing the evolution of grain size and morphology, along with the elemental analysis table (Ge, Sb, Te distribution across the film): (a) Film 1 (farthest from the source), (b) Film 2 (middle substrate), and (c) Film 3 (closest to the source). XRD patterns with reference cards from the PDF-2 database for (d) Film 1, (e) Film 2, and (f) Film 3. Optical spectra recorded from the center of the films in the 400–1700 nm range, with an inset showing the Tauc plot and the estimated optical band gap for (g) Film 1: $Ge_3Sb_2Te_6$, (h) Film 2: $Ge_2Sb_2Te_5$, and (i) Film 3: $GeSb_2Te_4$.

A decrease in plate size with increasing substrate–source distance can be explained by the temperature gradient along the CVD tube. At higher temperatures, enhanced diffusion promotes the growth of larger grains and more continuous films. In contrast, lower temperatures reduce atomic mobility, resulting in higher nucleation density and smaller grain sizes [31].

EDXS analysis confirmed a gradual variation in stoichiometry across the three films, consistent with the intended compositional gradient resulting from the linear substrate arrangement during CVD. As shown in Figure 1 (a-c), the atomic ratios indicate a transition from a Ge-rich composition (Film 1) to a Sb-rich composition (Film 3). Elemental distribution profiles are provided in Figure S2(b). Furthermore, XRD analysis revealed that the films crystallize in a hexagonal structure. Films 1 and Film 3 belong to the $R\bar{3}m$ space group, while Film 2 crystallizes in $P\bar{3}m1$. Based on the XRD and EDXS results, the film stoichiometries were determined as follows: Film 1 corresponds to $Ge_3Sb_2Te_6$, Film 2 to $Ge_2Sb_2Te_5$, and Film 3 to $GeSb_2Te_4$. Previous studies [23, 32, 33] have reported that temperature gradients and the specific distance between the source and substrates can induce compositional variations in the GeTe and GeTe-$Sb_2Te_3$ pseudo-binary systems. In agreement with these reports, CVD-grown films in the present study exhibit a compositional gradient, with $Ge_3Sb_2Te_6$, $Ge_2Sb_2Te_5$, and $GeSb_2Te_4$ phases forming as a function of the substrate-source distance along the tube. On the one hand, this film can be considered an in-plane gradient film. On the other hand, each composition is spatially separated on the substrate, exhibiting a distinct surface morphology, and can be studied independently, offering unique characteristics.

The absorption spectra and corresponding Tauc plots for each film are shown in Figure 1(g–i). All films exhibit good absorbance ($A$) in this spectral range, with an absorption minimum around 800 nm, followed by near-linear behavior from 1000 to 1700 nm. Nevertheless, some optical features depend on GST composition. Detailed optical characterization is provided in SI. The estimated optical band gaps, obtained by linear approximation of the Tauc plots, are 0.42, 0.49, and 0.55 eV for $GeSb_2Te_4$, $Ge_2Sb_2Te_5$, and $Ge_3Sb_2Te_6$, respectively. All films show a direct band gap. It was reported that within GeTe–$Sb_2Te_3$ pseudo-binary systems, the constituent compounds exhibit significantly different band gaps, with GeTe showing a relatively wide band gap (~0.66 eV), while $Sb_2Te_3$ is characterized by a much narrower band gap (~0.17 eV) [34]. Accordingly, an increase in the Ge/Sb atomic ratio results in a systematic widening of the band gap from approximately 0.5 eV to 0.61 eV (see Ref [34]), which has been attributed to the enhanced contribution of GeTe-related bonding states. In addition, similar trends have been observed in Ge-doped $Sb_2Te_3$, where an increase in Ge content leads to enlargement of the direct band gap [22]. Moreover, the grain size influences the band gap due to the spatial confinement of charge carriers. Specifically, a reduction in crystallite size leads to an increase in the band gap energy [35].

Within the GeTe–$Sb_2Te_3$ pseudo-binary line framework, the crystal structures of $Ge_3Sb_2Te_6$, $Ge_2Sb_2Te_5$, and $GeSb_2Te_4$ can be described as combinations of the GeTe and $Sb_2Te_3$ parent compounds [36, 37]. Therefore, it is reasonable to analyze the structural evolution from this perspective, where $GeSb_2Te_4$ corresponds to the GeTe–$Sb_2Te_3$ ternary system, $Ge_2Sb_2Te_5$ to 2GeTe–$Sb_2Te_3$, and $Ge_3Sb_2Te_6$ to 3GeTe–$Sb_2Te_3$. It is clearly observed that the amount of the GeTe insertions increases systematically across the compositions. Furthermore, the Raman spectral features may be interpreted as the combined contributions of these subsystems. Figure 2 shows the evolution of the crystalline structure from the substrate closest to the source (see Film 3, Figure 2(a)), through the intermediate substrate (see Film 2, Figure 2(b)), to the substrate farthest from the source (see Film 1, Figure 2(c)). The corresponding Raman spectra 1 to 9 reflect positions from the closest (1) to the farthest (9) region relative to the source, presenting three spectra for each film (two from the edges and one from the center). Optical microscope images of the measurement spots are also provided to visualize the quantitative changes in film surface morphology, clearly showing a gradual reduction in crystallite size as the distance from the substrate to the source decreases during deposition. Additionally, the Voight fitting was employed for a detailed structural analysis.

For a more reliable interpretation of the Raman spectra, reference measurements were performed on the $Ge_2Sb_2Te_5$ target crystal used for CVD, as well as on the parent compounds GeTe and $Sb_2Te_3$. Since the crystalline structures of GeTe and $Sb_2Te_3$ serve as base building blocks of the pseudo-binary compounds, their analysis is essential for understanding the structural features of these compounds. GeTe and $Sb_2Te_3$ crystals were synthesized by solid-state reaction from high-purity elements mixed in the required stoichiometric ratios and sealed in quartz ampoules. XRD confirmed phase purity (see Figure S3(a)). The fitted Raman spectra of these reference samples are also presented in Figure S3(b). Deconvolution of GeTe Raman spectrum resolves two main characteristic peaks at 86.8 $cm^{-1}$ (antisymmetric stretching mode, $E$) and 119.8 $cm^{-1}$ (symmetric stretching mode, $A_1$) [30, 38]. The $Sb_2Te_3$ crystal exhibits three prominent vibrational modes at approximately 69.6 $cm^{-1}$ (out-of-plane $A^1_{1g}$), 112.5 $cm^{-1}$ (in-plane $E^2_g$), and 167.0 $cm^{-1}$ (out-of-plane $A^2_{1g}$) [30]. The $Ge_2Sb_2Te_5$ target crystal closely resembles the single-crystalline hexagonal $Ge_2Sb_2Te_5$ reported in [39]. The Raman spectrum of $Ge_2Sb_2Te_5$ shows features similar to those of $Sb_2Te_3$, with some distinctions. In particular, the low-frequency peak at 55.5 $cm^{-1}$ is close to the $A^1_{1g}$ mode of $Sb_2Te_3$, while the high-frequency peak at 173.4 $cm^{-1}$ is comparable to the $A^2_{1g}$ mode. As reported in [39], this high-frequency mode originates from the $A^2_{1g}$ vibration of $Sb_2Te_3$. In contrast to the $Sb_2Te_3$ spectrum, a broad band in the range of 80–140 $cm^{-1}$ is observed in the $Ge_2Sb_2Te_5$ crystal, and this feature is attributed to lattice disorder induced by Ge/Sb site disorder [39]. Compositionally gradient films exhibit vibrational modes originating from both Ge-Te and Sb-Te subsystems. As a rule, these modes are attributed to the $GeTe_{4-n}Ge_n$ ($n$ = 0–3) tetrahedra and the $Sb_mTe_3$ ($m$ = 1, 2) structural units. Fitted peak positions and their assignments are summarized in Table 1. Among the Raman spectra of the GeTe-$Sb_2Te_3$ ternary system, the features most similar to those of the target crystal are observed for $GeSb_2Te_4$, which exhibits sharp low- and high-frequency modes; however, in the broad-band region, the peaks can be resolved. With increased GeTe content in the film composition, a gradual spectral transformation occurs, and for the $Ge_3Sb_2Te_6$ composition, the Raman spectra become similar to those of GeTe, showing two sharp dominant peaks. The vibrational modes of the $Sb_mTe_3$ ($m$ = 1, 2) structural units, highlighted in light-brown in Figure 2, are represented by $A^1_{1g}$ mode in the range 58.3 $cm^{-1}$ - 61.6 $cm^{-1}$, $E^2_g$ mode in the range 106.0 $cm^{-1}$ - 114.1 $cm^{-1}$, and the $A^2_{1g}$ mode in the range 172.5 $cm^{-1}$ - 184.2 $cm^{-1}$ [40, 41]. The vibrational modes of the $GeTe_{4-n}Ge_n$ ($n$ = 0–3) tetrahedra, highlighted in purple in Figure 2, are represented by $E$ mode in the range 63.1 $cm^{-1}$ – 74.1 $cm^{-1}$, $A_1$ corner-sharing mode in the range 124.7 $cm^{-1}$ – 146.3 $cm^{-1}$, and the $A_1$ edge-sharing mode in the range 142.9 $cm^{-1}$ – 150.1 $cm^{-1}$ [40-42]. Additionally, two peaks are observed (highlighted in gray in Figure 2): one in the 86 $cm^{-1}$ –96 $cm^{-1}$ range and another in the 161 $cm^{-1}$ – 166 $cm^{-1}$ range. They most likely arise from structural defects, which are abundant in GST pseudo-binary compounds [43]. These bands have been attributed either to the threefold-coordinated Te Raman mode [44, 45] or to defective octahedral Ge [46, 47]. In particular, bands at 156 $cm^{-1}$ and 95 $cm^{-1}$ were assigned to defective octahedral Ge in the crystalline GeTe phase change material [54]. Furthermore, a peak at 162 $cm^{-1}$ was linked to overlapping Ge–Te and Te–Te contributions in CVD-grown GeTe films [43]. Another study attributed the 90 $cm^{-1}$ peak in Ge-rich GST films to combined contributions from Te–Te stretching and Ge–Te stretching modes [44].

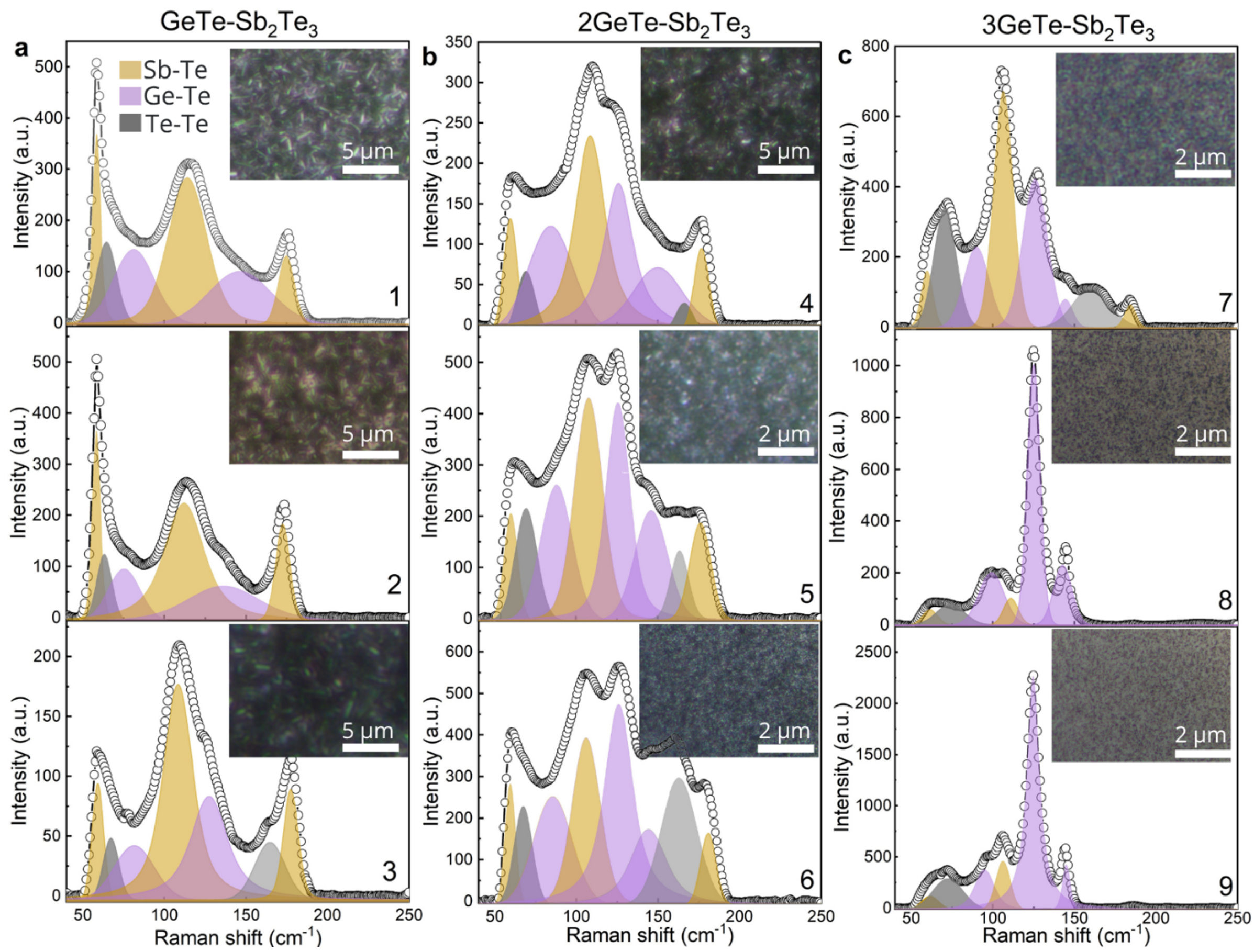

Figure 2. (a) Raman spectra of as-deposited films recorded sequentially from the region closest to the source to the farthest point across the surfaces of Film 1, Film 2, and Film 3, showing the gradual evolution of structural features along the compositional gradient. Optical microscopy images of the corresponding measurement spots are included to visualize quantitative changes in surface morphology. Colors indicate the following vibrational modes: yellow for $Sb_mTe_3$ ($m$ = 1, 2) subsystem, purple for $GeTe_{4-n}Ge_n$ ($n$ = 0–3) tetrahedra, and gray for Te-Te- or defective octahedral Ge-related vibrations.

Table 1. Voigt-function deconvolution of Raman spectra for each composition, along with the corresponding band assignments.

| Spectra No. | Subsystem | $Sb_mTe_3$ ($m$ = 1, 2) | | | $GeTe_{4-n}Ge_n$ ($n$ = 0–3) | | | Te- or defective octahedral Ge | |
|---|---|---|---|---|---|---|---|---|---|
| | Band assignment | $A^1_{1g}$ | $E^2_g$ | $A^2_{1g}$ | E | $A_1$ corner-sharing | $A_1$ edge-sharing | | |
| | Composition | Peak position, $cm^{-1}$ | | | | | | | |
| 1 | $GeSb_2Te_4$ | 58.4 | 114.1 | 174.8 | 64.5 | 146.3 | - | 81.2 | - |
| 2 | | 58.3 | 112.1 | 172.5 | 63.1 | 136.1 | - | 75.3 | - |
| 3 | | 59.2 | 108.4 | 177.5 | 67.2 | 127.4 | - | 81.5 | 164.7 |
| 4 | $Ge_2Sb_2Te_5$ | 59.7 | 108.7 | 179.6 | 69.3 | 126.1 | 150.1 | 84.5 | 166.4 |
| 5 | | 60.0 | 107.7 | 176.2 | 69.4 | 125.7 | 146.1 | 88.0 | 163.5 |

| 6 | | 59.4 | 106.0 | 180.9 | 67.4 | 125.9 | 144.2 | 85.5 | 162.9 |
|---|---|---|---|---|---|---|---|---|---|
| 7 | $Ge_3Sb_2Te_6$ | 59.7 | 106.4 | 184.2 | 70.7 | 125.9 | 144.4 | 90.0 | 160.6 |
| 8 | | 61.5 | 111.1 | - | 74.1 | 125.1 | 142.9 | 99.2 | - |
| 9 | | 61.6 | 106.2 | - | 72.5 | 124.7 | 145.0 | 94.9 | - |

In $GeSb_2Te_4$, $Sb_mTe_3$ ($m$ = 1, 2) modes appear as sharp, intense peaks, while the intensity of vibrational modes of the $GeTe_{4-n}Ge_n$ ($n$ = 0–3) tetrahedra is lower. It was reported before that Sb–Te vibrations dominate over GeTe vibrations because of the latter's lower polarizability [48]. In $Ge_2Sb_2Te_5$, the contributions of both subsystems become comparable. As the GeTe content increases, its contribution grows, balancing the $Sb_mTe_3$ ($m$ = 1, 2) modes. In $Ge_3Sb_2Te_6$, the $Sb_mTe_3$ ($m$ = 1, 2) modes nearly disappear, being suppressed by the dominant vibrational contributions of the $GeTe_{4-n}Ge_n$ ($n$ = 0–3) tetrahedra. The $A_1$ corner-sharing vibration becomes the most intense and sharp peak, indicating the dominant contribution of $GeTe_{4-n}Ge_n$ ($n$ = 0–3) in this composition. Regarding the Te- and/or defective octahedral Ge-related modes, their contribution is present in all compositions, especially in spectra collected at film edges. In $Ge_2Sb_2Te_5$ (particularly in spectra 5 and 6), these modes are the most intense, suggesting that this composition is more structurally disordered compared with $GeSb_2Te_4$ and $Ge_3Sb_2Te_6$ compositions.

With increasing GeTe content in the compounds, the positions of the Raman peaks undergo systematic shifts. In particular, the vibrational modes associated with $GeTe_{4-n}Ge_n$ (n = 0–3) tetrahedra are the most sensitive. With increasing GeTe content, the $E$ mode shifts toward lower wavenumbers (red shift), while the $A$ modes shift toward higher wavenumbers (blue shift), both with an overall magnitude of approximately 11 $cm^{-1}$ (between spectra 2 and 8 in Figure 2). In contrast, the $Sb_mTe_3$-related modes exhibit less pronounced but still noticeable shifts. Notably, the $A^2_{1g}$ mode disappears entirely in the $Ge_3Sb_2Te_6$ composition (see spectra 8 and 9 in Figure 2), indicating a significant modification of the local structural environment. These peak shifts can be attributed not only to compositional variations but also to changes in crystallite size (see insets in Figure 2) and film thickness. It has been previously demonstrated that a reduction in crystallite size leads to a downshift of phonon modes in GeTe [49]. In addition, a strong correlation between the thickness of crystalline GeTe films and their vibrational characteristics has been reported, where both $A$ and $E$ modes decrease in frequency with increasing film thickness [30]. In the present case, a similar effect is observed: the $A$ modes decrease in frequency with increasing thickness, while the $E$ mode position downshifts in response to a reduction in crystallite size.

## 2.2 Optical Behavior of In-Plane Gradient Ge–Sb–Te Films

The 400-1000 nm spectral region showed a pronounced absorption minimum ($\lambda_{min}$) for all three compositions, whereas in the 1000-1700 nm region, not all films exhibit sharp spectral features (see Figure 1(g-i)). Moreover, considering the plate thicknesses for all three films (0.2 µm for $Ge_3Sb_2Te_6$, 0.3 µm for $Ge_2Sb_2Te_5$, and 0.4 µm for $GeSb_2Te_4$), the visible spectral range presents the most informative window for establishing structure-properties relationships. The $L$-direction measurements were expected to reflect changes in film composition, thickness, and morphology, while the $T$-direction measurements were expected to exhibit nearly constant composition and thickness. To quantitatively characterize the spatial anisotropy of the as-deposited films, the position of the absorption minimum $\lambda_{min}$ and its intensity $A_{min}$ were plotted as functions of the distance where spectra were collected from one edge to the opposite (9 measurement points): for $GeSb_2Te_4$ (see Figure 3 (a-d)), for $Ge_2Sb_2Te_5$ (see

Figure 3(e-h)), and for $Ge_3Sb_2Te_6$ (see Figure 3 (i-l)). The axis scales were maintained within identical ranges across all plots to enable a consistent and more rigorous visual comparison of the observed differences.

In the 400–1000 nm spectral range, the absorption spectra of all compositions exhibit similar behavior, characterized by a gradual decrease in absorbance reaching a minimum around 800–900 nm, followed by a subsequent increase. As the measurement position shifts from one edge of the film to the other, both along the composition gradient and perpendicular to it, a gradual increase in the wavelength of the absorption minimum, accompanied by a change in its intensity, is observed. Lower absorbance values are observed for the $GeSb_2Te_4$ (see Figure 3(a, b)) and $Ge_2Sb_2Te_5$ (see Figure 3 (e, f)) films, with a corresponding spread in absorption intensity significantly smaller. $Ge_3Sb_2Te_6$ film demonstrates the highest absorbance values (see Figure 3 (i, j)). Notably, the absorption signals collected from the film edges reach saturation at $A > 1$, indicating that the measured intensity exceeds the reliable detection limit, likely due to the high density of defects formed in these regions. Overall, moving along the deposition direction from $GeSb_2Te_4$ film to $Ge_3Sb_2Te_6$ film, the absorption minimum monotonically shifts from 815.9 nm to 889.8 nm, accompanied by a 5.9-fold change in the absorption intensity at the minimum point. The transition from $GeSb_2Te_4$ to $Ge_3Sb_2Te_6$ is accompanied by a simultaneous increase in GeTe structural units, film thickness, and surface morphology evolution. According to Raman analysis, the vibrational modes associated with $GeTe_{4-n}Ge_n$ ($n$ = 0–3) tetrahedra are the most sensitive to this compositional transition. At the same time, the surface morphology evolves from anisotropic plate-like crystallites to smaller anisotropic crystallites in $Ge_2Sb_2Te_5$ and finally to fine, nearly isotropic crystallites in $Ge_3Sb_2Te_6$. With increasing GeTe-related structural contribution, grain boundary densities, and film thicknesses along the $L$-direction, $\lambda_{min}$ gradually shifts toward longer wavelengths and $A_{min}$ increases within the 400-1000 nm spectral range, indicating a combined influence of these factors on the optical response.

At the same time, noticeable variations in the optical properties are also observed within each composition. Notably, $Ge_3Sb_2Te_6$ film exhibits the largest variation in absorption intensity across this spectral range, in both the $L$- and $T$-directions. For $Ge_2Sb_2Te_5$ and $GeSb_2Te_4$ films, along both directions, the position and intensity of the absorption minimum decrease monotonically from one edge to the other (see Figure 3(c, d, g, h)), and along the $T$-direction the variations are minimal. The corresponding variations in the position and intensity of the absorption minima are summarized in Table 2.

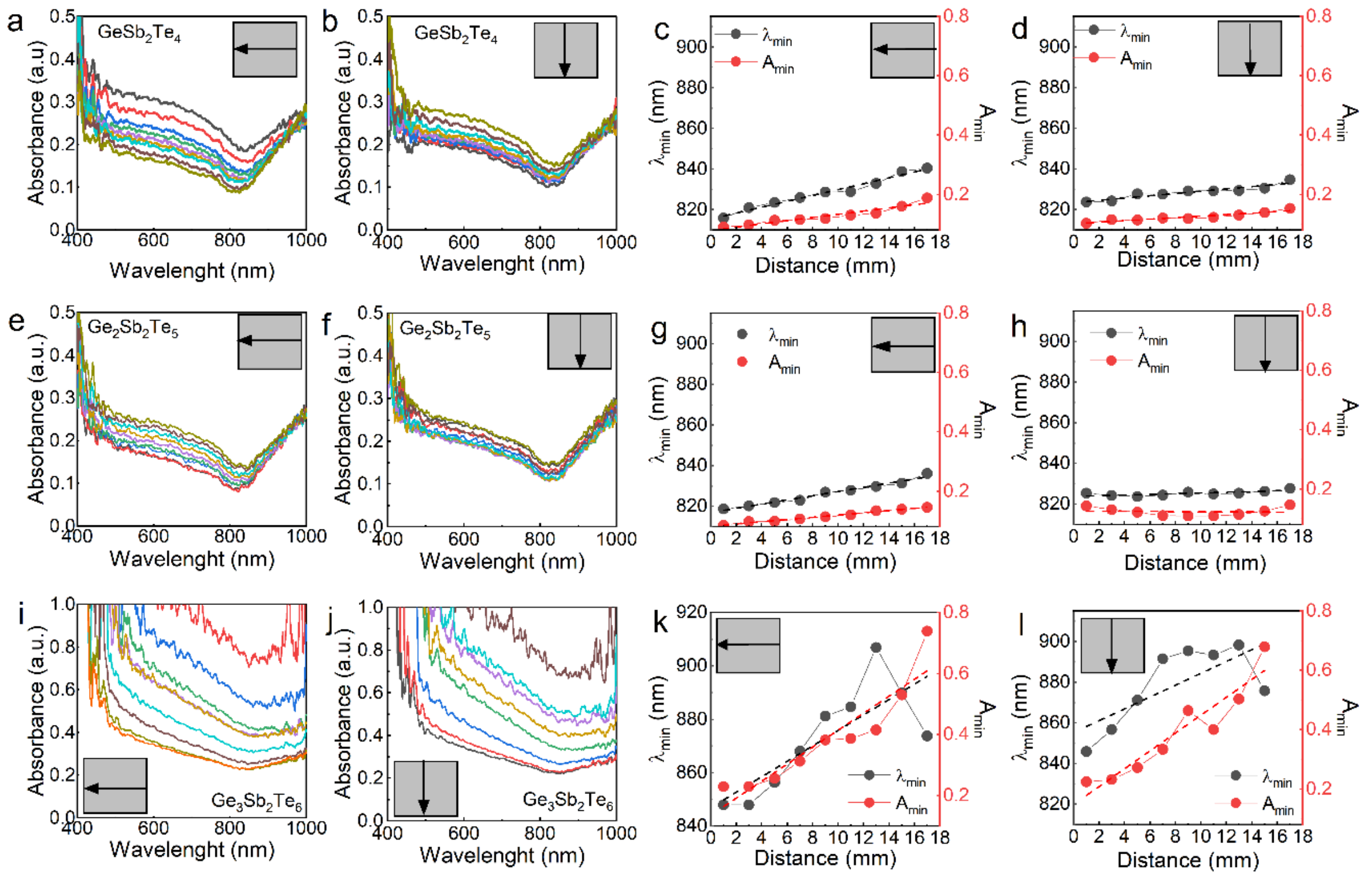

Figure 3. Absorbance spectra of GST films measured in the 400-1000 nm range. Absorbance spectra of $GeSb_2Te_4$ measured along two directions: (a) *L*-direction and (b) *T*-direction. The corresponding dependences of the absorbance minimum position and the absorbance intensity at this minimum along the $GeSb_2Te_4$ film are shown for the (c) *L*-direction and (d) *T*-direction. Absorbance spectra of $Ge_2Sb_2Te_5$ measured along the (e) *L*-direction and (f) *T*-direction. The corresponding dependences of the absorbance minimum position and the absorbance intensity at this minimum along the $Ge_2Sb_2Te_5$ film for the (g) *L*-direction and (h) *T*-direction. Absorbance spectra of $Ge_3Sb_2Te_6$ measured along the (i) *L*-direction and (j) *T*-direction. The corresponding dependences of the absorbance minimum position and the absorbance intensity at this minimum along the $Ge_3Sb_2Te_6$ film for the (k) *L*-direction, (l) *T*-direction. A schematic illustration of the measurement direction is provided with all graphs. Measurement step: 2 mm. Dashed lines represent the linear approximation of the obtained dependencies.

Linear regression was performed separately for the $\lambda_{min}$ and $A_{min}$ variations along the *L*- and *T*-directions, yielding slope parameters for each film: $S_\lambda^L$, $S_A^L$ $S_\lambda^T$, and $S_A^T$. These slopes represent the rate of change of the optical absorption along the film. $S_\lambda$ reflects the spatial gradient in the effective optical thickness $(n \cdot d)$, while $S_A$ quantifies the spatial evolution of the extinction intensity at the minimum. The directional contrast in these spatial gradients is then determined by the differences $\Delta S_\lambda = |S_\lambda^L - S_\lambda^T|$ and $\Delta S_A = |S_A^L - S_A^T|$. A high $\Delta S$ indicates that the *L*- and *T*-directions exhibit markedly different spatial rates of change, i.e., the effective optical thickness evolves differently when probed along the deposition direction versus perpendicular to it. Such a situation arises when the crystalline plates have a preferred orientation or when their packing density varies strongly with direction. Conversely, a low $\Delta S$ (close to zero) value means that the spatial gradients are nearly identical in both directions, implying that the film is effectively isotropic in terms of its gradient structure. Comparing $\Delta S_\lambda$ and $\Delta S_A$ across the three films thus provides a quantitative measure of how the optical absorption correlates with the microstructural features.

Table 2. Shifts in $\lambda_{min}$ and corresponding absorption intensity ($\Delta A_{min}$), measured along the *L*- and *T*-directions, together with the derived rates of optical absorption change ($S_\lambda^L$, $S_A^L$ $S_\lambda^T$, and $S_A^T$) and the corresponding spatial variation parameters $\Delta S_\lambda$ and $\Delta S_A$ across the in-plane compositionally gradient Ge–Sb–Te films.

| Film | *L*-direction | *T*-direction | | |
|---|---|---|---|---|

| | $\Delta\lambda_{min}$, nm | $\Delta A_{min}$ | $S_{\lambda}^{L}$, | $S_{A}^{L}$, $10^{-3}$ | $\Delta\lambda_{min}$, nm | $\Delta A_{min}$ | $S_{\lambda}^{T}$, | $S_{A}^{T}$, $10^{-3}$ | $\Delta S_{\lambda}$ | $\Delta S_{A}$, $10^{-3}$ |
|---|---|---|---|---|---|---|---|---|---|---|
| $GeSb_2Te_4$ | 24.5 | 0.10 | 1.44 | 5.38 | 10.9 | 0.05 | 0.56 | 2.58 | 0.88 | 2.8 |
| $Ge_2Sb_2Te_5$ | 17.4 | 0.06 | 1.03 | 3.78 | 2.5 | 0.004 | 0.16 | -0.15 | 0.87 | 3.6 |
| $Ge_3Sb_2Te_6$ | 26.0 | 0.51 | 2.89 | 27.8 | 29.9 | 0.46 | 2.91 | 30.1 | 0.02 | 2.3 |

Within the individual films, for $Ge_2Sb_2Te_5$ and $GeSb_2Te_4$ compositions, the variations along the *T*-direction are systematically smaller than those along the $L$-direction ($S_{\lambda(A)}^{L}$>$S_{\lambda(A)}^{T}$), consistent with the absence of a compositional or thickness gradient and indicating a reduced contribution from directional variations in morphology. It is worth noting the formation of anisotropic plate-like crystallites contributing to the film surface morphology, which are preferentially oriented perpendicular to the substrate surface (see Figure 1(b, c)). $Ge_2Sb_2Te_5$ exhibits well-defined plate-like structures with an aspect ratio (length/thickness) of ~4.7, whereas $GeSb_2Te_4$ is characterized by more developed plate-like structures with an aspect ratio of ~9.0. Therefore, the observed differences in the *T*-direction may be associated with variations in the plate morphology and orientation relative to the incident light, resulting in lower $S_{\lambda}^{T}$ values. Notably, despite its lower structural disorder, $GeSb_2Te_4$ exhibits a higher $S_{\lambda}^{T}$ value (0.56 compared with 0.16 for $Ge_2Sb_2Te_5$), suggesting that the more pronounced platelet morphology may have a stronger influence on the directional optical response. $Ge_2Sb_2Te_5$ exhibits the lowest $S_{\lambda}^{L(T)}$and $S_{A}^{L(T)}$parameters among the investigated films. Raman analysis reveals comparable contributions from Ge-Te and Sb-Te related vibrational modes together with pronounced Te-Te- or Ge-related defective octahedral modes, indicating the higher degree of local structural heterogeneity [39]. Overall, the resulting spatial variations for $Ge_2Sb_2Te_5$ and $GeSb_2Te_4$ films remain comparable, with $\Delta S_{\lambda}$values close to ~0.9.

For $Ge_3Sb_2Te_6$ $S_{\lambda}^{L(T)}$ and $S_{A}^{L(T)}$parameters are significantly higher compared with other compositions. This film is characterized by dominant $GeTe_{4-n}Ge_n$ ($n$ = 0–3) tetrahedra-related modes and a higher density of grain boundaries. These microstructural features may contribute to enhanced local variations in the optical response, resulting in larger spatial gradients. Although the film thicknesses and composition are considered nearly constant along the *T*-direction, the optical variations remain comparable to those observed along the $L$-direction, leading to relatively small $\Delta S_{\lambda(A)}$. This indicates that the optical response of $Ge_3Sb_2Te_6$ is influenced not only by the compositional gradient but also by morphology-related effects, which may contribute more significantly to the optical response of $Ge_3Sb_2Te_6$ than in other compositions. Although the individual contributions of these factors cannot be quantitatively separated in the present study.

Overall, in-plane gradient film containing $GeSb_2Te_4$, $Ge_2Sb_2Te_5$ and $Ge_3Sb_2Te_6$ compositions exhibits a unified optical behavior in the 400–1000 nm range, indicating a common underlying physical origin. At the same time, the system represents an interesting platform for selective light absorption, which is governed not only by the chemical composition of the films but also by their surface morphology.

**2.3 Local Laser-Controlled Modification of the Optical Response**

In phase-change Ge–Sb–Te systems, laser irradiation is known to induce structural transformations that strongly affect optical properties [40, 50]. Previous studies of laser-induced phase transitions in GST films have predominantly employed single-pulse nanosecond or femtosecond irradiation with fluences in the range of 50–250 mJ/cm$^2$, typically for thin films (≤100 nm) [24, 51]. Under these conditions, the absorbed energy density is sufficient to induce melting across the full film thickness, followed by rapid quenching into the amorphous state. In the present work, the investigated films are significantly thicker. At Vis and NIR wavelengths, the optical absorption depth in crystalline GST is typically limited to several tens of nanometers. As a result, single-pulse excitation primarily deposits energy within a surface layer, while the underlying volume remains largely unaffected. This thickness disparity naturally shifts the relevant laser–matter interaction regime from single-pulse melting toward cumulative excitation processes.

The optical properties after laser irradiation are presented in Figure 4.

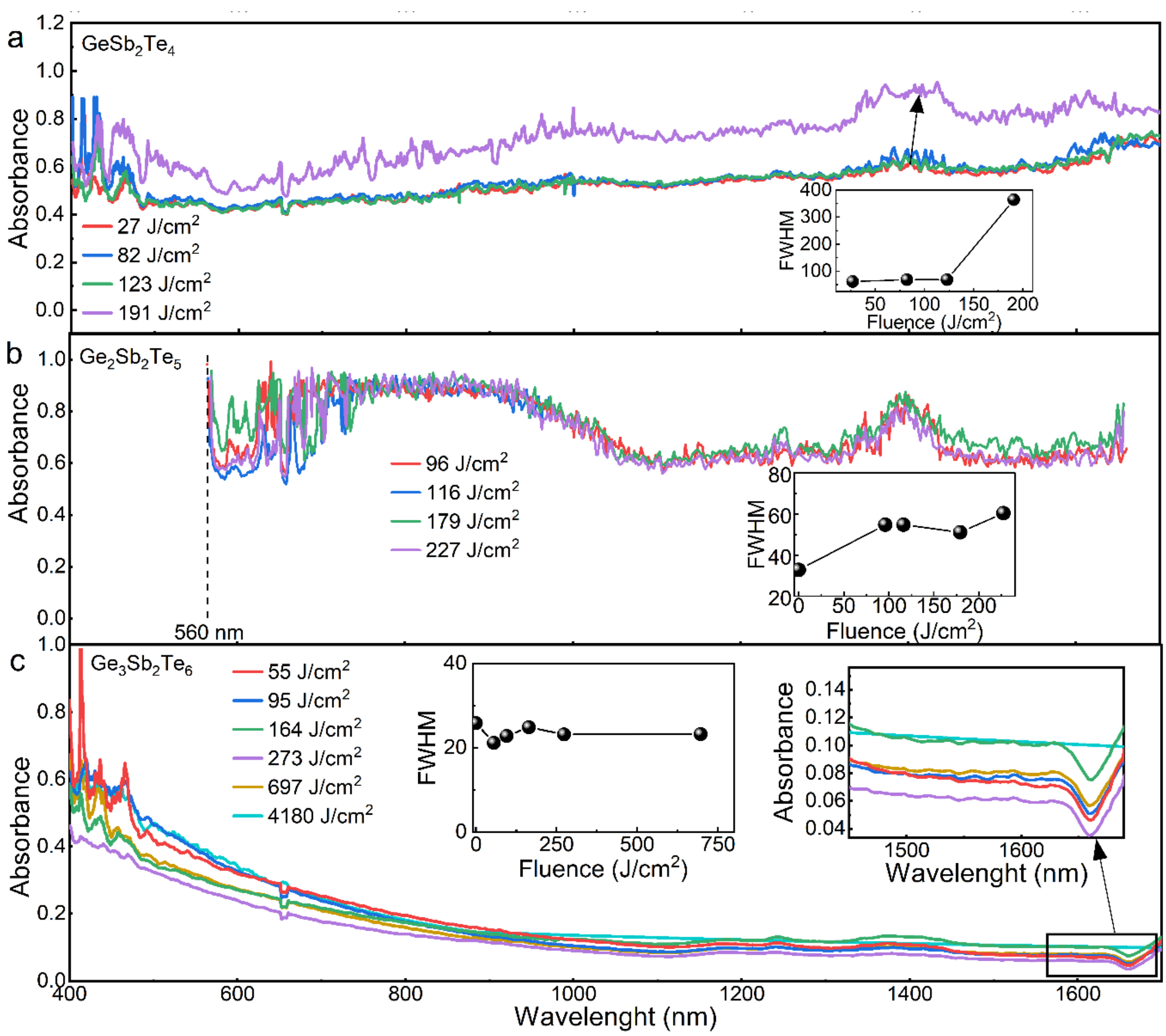

Figure 4. Optical properties of Ge–Sb–Te films measured after laser irradiation at various fluences in the 400-1700 nm spectral range: (a) $GeSb_2Te_4$ film, with the inset showing the dependence of the FWHM of the ~1400 nm peak on fluence, (b) $Ge_2Sb_2Te_5$ film, with the inset showing the dependence of the FWHM of the ~1400 nm peak on fluence (c) $Ge_3Sb_2Te_6$ film, with a zoomed-in view of the absorption peak at 1660 nm and the inset showing the dependence of the peak FWHM on fluence.

For all films, even the lowest applied laser fluence induced measurable modifications in the optical response within the corresponding spectral range. A detailed description of the optical modifications after laser irradiation is provided in SI. In the 400-1000 nm spectral range, all films show a pronounced deviation from their initial optical response. In particular, $\lambda_{min}$ around 800 nm, characteristic of all samples before irradiation, disappears. In the 1000-1700 nm range, the $GeSb_2Te_4$ film demonstrates high sensitivity to fluence, exhibiting fluence-dependent peak broadening. The $Ge_2Sb_2Te_5$ film largely preserves its absorbance after irradiation, while the absorption peaks become more pronounced. Similarly, the $Ge_3Sb_2Te_6$ film exhibits a more pronounced absorption minimum and undergoes only minor variations even at elevated fluence values. To quantitatively illustrate these changes, the absorption contrast was evaluated at the absorption minimum or maximum in the Vis and NIR spectral ranges. The absorbance contrast was estimated as $A_i/A_0$, where $A_i$ represents the absorbance following laser irradiation at the corresponding fluence, and $A_0$ is the initial absorbance of the as-deposited film. The corresponding values are presented in Table 3.

Table 3. Absorbance contrast of Ge–Sb–Te films following laser irradiation at various fluence levels.

<table>
<tr><th colspan="3">$Ge_3Sb_2Te_6$</th><th colspan="3">$Ge_2Sb_2Te_5$</th><th colspan="3">$GeSb_2Te_4$</th></tr>
<tr><th rowspan="2">Fluence, J/cm$^2$</th><th colspan="2">$A_i/A_0$</th><th rowspan="2">Fluence, J/cm$^2$</th><th colspan="2">$A_i/A_0$</th><th rowspan="2">Fluence, J/cm$^2$</th><th colspan="2">$A_i/A_0$</th></tr>
<tr><th>VIS at ~881 nm</th><th>IR at ~1660 nm</th><th>VIS at ~827 nm</th><th>IR at ~1400 nm</th><th>VIS at ~829 nm</th><th>IR at ~1400 nm</th></tr>
<tr><td>55</td><td>0.34</td><td>0.14</td><td>96</td><td>7.7</td><td>1.7</td><td>27</td><td>4.0</td><td>2.1</td></tr>
<tr><td>95</td><td>0.36</td><td>0.13</td><td>116</td><td>7.8</td><td>1.7</td><td>82</td><td>3.9</td><td>2.2</td></tr>
<tr><td>164</td><td>0.38</td><td>0.19</td><td rowspan="2">179</td><td rowspan="2">8.1</td><td rowspan="2">1.7</td><td rowspan="2">123</td><td rowspan="2">3.9</td><td rowspan="2">2.2</td></tr>
<tr><td>273</td><td>0.32</td><td>0.09</td></tr>
<tr><td>697</td><td>0.42</td><td>0.12</td><td rowspan="2">227</td><td rowspan="2">8.1</td><td rowspan="2">1.5</td><td rowspan="2">191</td><td rowspan="2">5.6</td><td rowspan="2">3.2</td></tr>
<tr><td>4180</td><td>0.37</td><td>0.12</td></tr>
</table>

The pronounced decrease in absorbance is observed for the $Ge_3Sb_2Te_6$ film, where the absorbance in both the Vis and NIR regions is reduced severalfold after laser irradiation, with a 2.4-3.1-fold decrease in the Vis region and a 5.3-11.1-fold decrease in the NIR region. In contrast, the $Ge_2Sb_2Te_5$ and $GeSb_2Te_4$ films exhibit a significant enhancement of absorbance, particularly in Vis range. For $Ge_2Sb_2Te_5$, the absorbance increases by nearly an order of magnitude in Vis region, while NIR absorbance changes in a more moderate manner. The $GeSb_2Te_4$ film also demonstrates enhanced absorbance in both spectral regions, with a more pronounced increase observed at higher laser fluences.

### 2.4. Structure–Property Relationships in Laser-Irradiated Spots

It is a well-known fact that light irradiation promotes defect formation and alters the structural and electronic properties of the materials [52]. Specifically, in crystalline systems, illumination can induce the generation of vacancies, bond breakage, and various defect states, such as dangling bonds and valence-alternation pairs. The underlying mechanisms involve photon-induced excitation of electron–hole pairs, which leads to bond reconfiguration and the creation or migration of defect centers [52]. Thus, light affects crystalline materials by facilitating the formation and redistribution of defects, thereby modifying their electronic properties and structural stability. The initial features of the surface morphology of the obtained films should also be considered, as they have been shown to significantly influence the optical behavior of the films in the 400–1000 nm spectral range. Understanding these processes is crucial for the development of devices based on crystalline GST materials, such as optical and photonic components, since light-induced defect formation can both enhance and degrade their functional performance.

To investigate structural changes induced by multiple femtosecond laser pulses, additional spectroscopic and microscopic analyses were performed. Using the $Ge_3Sb_2Te_6$ film, a comprehensive structural characterization was carried out over a wide range of laser fluences. This composition was selected because its surface morphology consists predominantly of relatively isotropic features with uniform shape and size, enabling clearer identification and interpretation of laser-induced microstructural transformations. In contrast, the $GeSb_2Te_4$ and $Ge_2Sb_2Te_5$ films exhibit more anisotropic, platelet-like structures, which complicate the analysis of irradiation-induced morphological evolution.

The highest fluence of 4180 J/cm$^2$ produced the most pronounced modification, clearly observable under an optical microscope. Upon closer inspection of the irradiated area (see Figure 5(a)), three distinct zones can be identified, extending from the center toward the edge. In the employed femtosecond laser system, the spatial distribution of the beam is well described by a Gaussian profile [53]. The spatial distribution of fluence (or intensity), in this case, as a function of distance from the beam center, can be expressed as:

$$F(r) = F_{peak} \exp\left(\frac{-2r^2}{w_0^2}\right), \tag{1}$$

where $F(r)$ is the fluence at a given radial position $r$, $F_{peak}$ is the maximum fluence at the center of the beam, $w_0$ is the beam waist (defined as the radius at which the intensity decreases to $1/e^2$ of its maximum value [54]). At the beam center ($r = 0$), the fluence reaches its maximum value $F_{peak}$, while at $r = w_0$ it decreases to approximately $F_{peak} \cdot e^{-2} \approx 0.135 F_{peak}$, demonstrating the strong spatial confinement of energy in the central region of the beam. The average fluence over the entire beam cross-section is given by $F_{avg} = \frac{E}{\pi w_0^2}$, where $E$ is the pulse energy. For a Gaussian beam, the peak fluence is related to the average fluence by $F_{peak} = 2F_{avg}$. This non-uniform energy distribution has significant experimental consequences, as the central region of the irradiated spot experiences a significantly higher fluence compared to the periphery, often leading to spatially resolved structural modifications. As a result, concentric zones with different structural or phase states can form, depending on local fluence thresholds within the material.

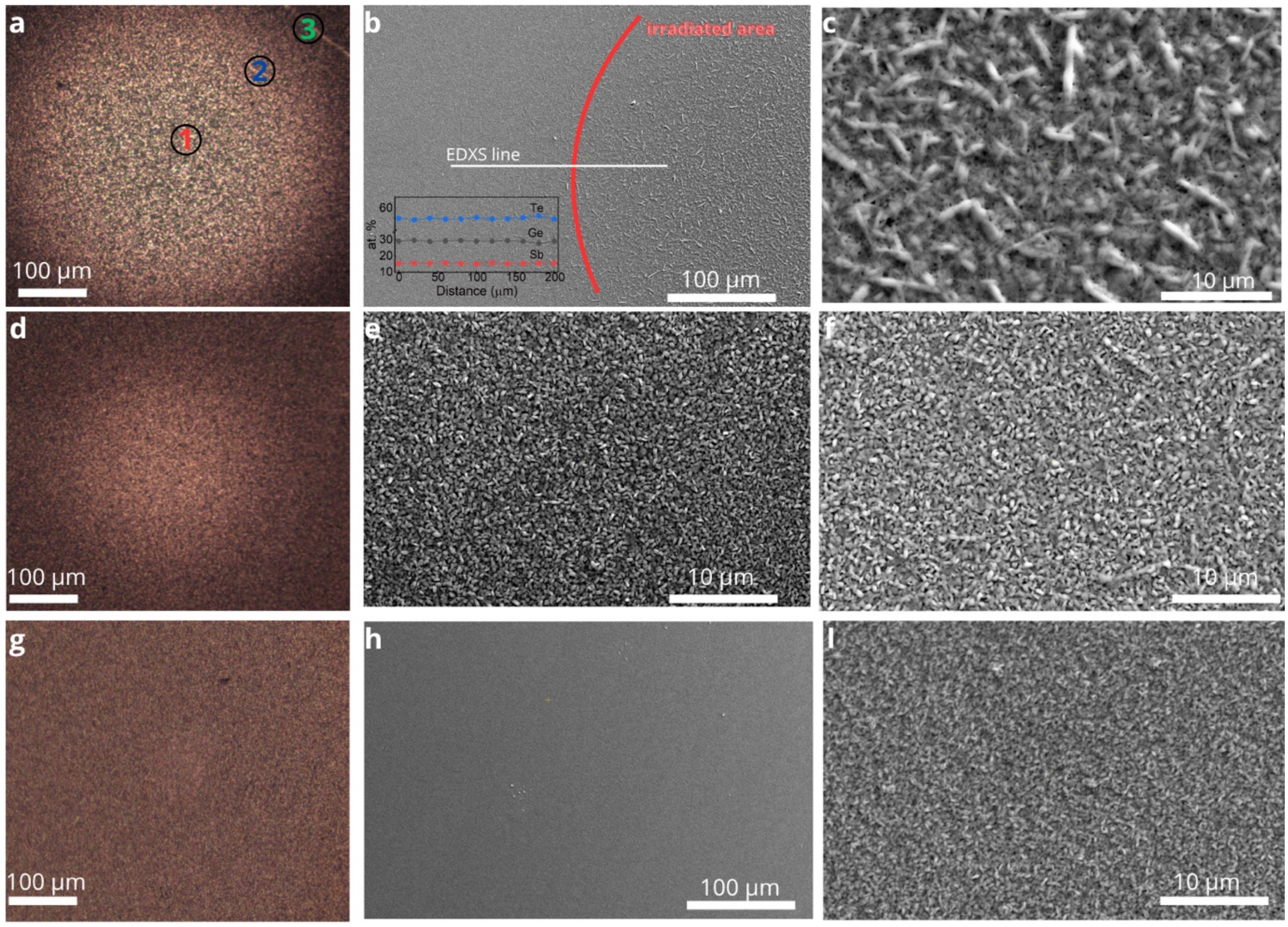

Figure 5. Microstructural analysis of multiple femtosecond laser-irradiated regions in $Ge_3Sb_2Te_6$ film. Microstructural analysis of the laser-irradiated region at 4180 J/cm$^2$, where (a) an optical microscopy image of the irradiated area, showing three visually distinct zones (1-3), (b) a SEM image of the transition boundary marking the onset of the irradiated region, together with a line-profile EDXS analysis, and (c) a SEM image of the microstructure at the center of the irradiated spot. Microstructural analysis of the laser-irradiated region at 697 J/cm$^2$, where (d) an optical image of the irradiated area, a SEM image of the microstructure (e) at the edge and (f) at the center of the irradiated spot. Microstructural analysis of the laser-irradiated region at 273 J/cm$^2$, where (g) an optical image of the irradiated area, and (h-i) SEM images of the corresponding microstructure within the irradiated region.

So, for $F_{avg}$ =4180 J/cm$^2$, the peak fluence $F_{peak}$ is 8360 J/cm$^2$. A more detailed examination of the irradiated region (see Figure 5(b, c)) reveals changes in surface morphology, specifically the melting of crystalline plates. The melted plates form narrow rod-like structures with an average length of 3.8 µm, while some of them reach up to 10 µm. To determine whether any changes in elemental composition occur within the irradiated zone, EDXS line-scan analysis was performed across a region encompassing both the non-irradiated and irradiated areas. The results show that the elemental composition remains unchanged, indicating that only laser-induced plate fusion takes place. Additional SEM images of the laser-irradiated region at 4180 J/cm$^2$ are provided in Figure S4.

The radial gradient for the irradiated spot at 697 J/cm$^2$ is also observed (see Figure 5d), with a corresponding peak fluence of $F_{peak}$ = 1393 J/cm$^2$. A comparison of the regions at the edge and center of the irradiated area (see Figure 5e, f) reveals a distinct contrast associated with plates melting in the film. At the edge, rod-like structures are not observed, and the average plate size is approximately 0.6 µm. In contrast, at the center, the formed rod-like structures exhibit an average length of 1.8 µm, with the longest rods reaching up to 5 µm. The microscopic analysis of the area irradiated at 273 J/cm$^2$ (see Figure 5(g, i)) did not reveal any discernible changes. Both optical and SEM imaging failed to identify any visible modifications that could be reliably detected within the resolution limits of the techniques employed. This indicates that the SEM analysis of regions exposed to lower fluences was not informative, as no detectable changes could be identified.

spectroscopy (see Figure 6(a-h)). The non-irradiated film spectrum was dominated by GeTe-related vibrational modes (see Figure 2, spectrum 8), reflecting the initial structural state. Upon irradiation, systematic changes in the Raman spectra were observed. The most pronounced evolution occurred for the Te-Te- or octahedral Ge- related modes near 99.2 $cm^{-1}$, which shifted toward lower wavenumbers, while the mode near 163.1 $cm^{-1}$ shifted toward higher wavenumbers with increasing fluence. In contrast, the modes associated with the $Sb_mTe_3$ ($m$=1, 2) subsystem exhibited only minor position changes, although the $E^1_{1g}$ mode became detectable at fluences above 697 J/$cm^2$. The $GeTe_{4-n}Ge_n$ ($n$ = 0–3) tetrahedra-related $A_1$ corner- and edge-sharing modes exhibited a shift toward higher wavenumbers, whereas the $E$ mode showed a slight shift toward lower wavenumbers with increasing fluence. Changes in the peak widths further confirmed increasing material imperfections due to laser-induced structural modifications [55]. The $E^2_{1g}$ mode of the $Sb_mTe_3$ ($m$=1, 2) subsystem exhibited substantial broadening, whereas the $A^1_{1g}$ mode became narrower. Similarly, the $GeTe_{4-n}Ge_n$ ($n$ = 0–3) tetrahedral modes showed a contrasting evolution, with the $E$ mode narrowing and the $A_1$ modes broadening. The mode near 99.2 $cm^{-1}$ also exhibited broadening with increasing fluence. At the highest irradiation dose (4180 J/ $cm^2$), Raman spectra collected from different positions within the laser spot revealed a gradual shift of most modes toward lower wavenumbers and an overall increase in peak broadening toward the spot center, where the local energy density is the highest (see Figure S5).

Taken together, starting from a fluence of 55 J/$cm^2$, Raman spectroscopy reveals structural rearrangements, evidenced by peak broadening and shifts in band positions, increasing contribution of defective modes. These modifications become increasingly pronounced toward the center of the irradiated region, where the deposited energy density is highest. At fluences of 697 J/$cm^2$ and above, the formation of rod-like structures is observed by SEM, indicating concurrent morphological evolution. The observed optical changes, therefore, correlate with both structural and morphological modifications induced by laser irradiation, suggesting that both factors contribute to the resulting optical response of $Ge_3Sb_2Te_6$.

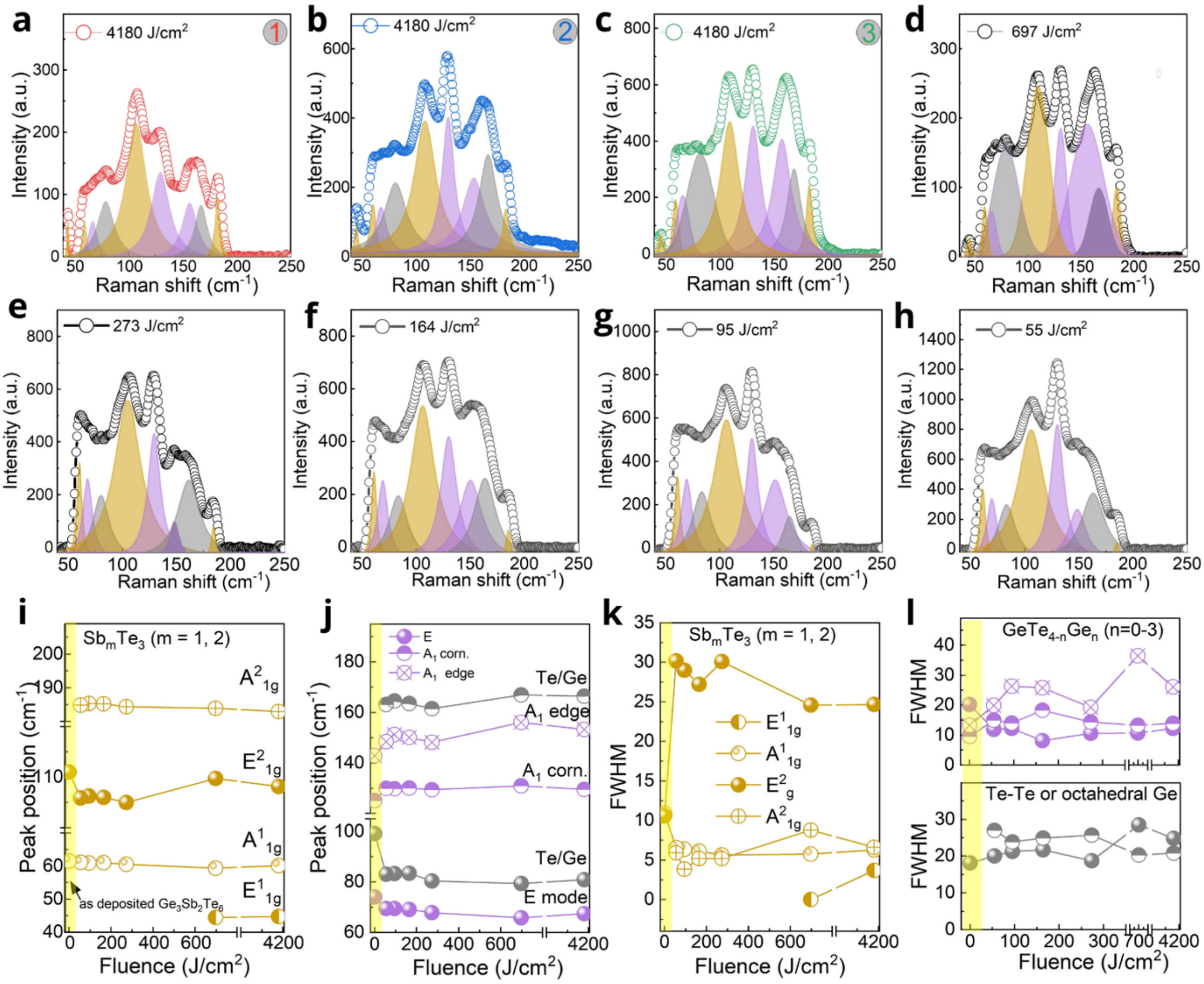

Figure 6. Raman spectra of $Ge_3Sb_2Te_6$ film after multiple femtosecond laser irradiation. Spectra acquired after a fluence of 4180 J/cm$^2$ are shown for different spots: (a) the center, (b) the edge, and (c) the boundary with the non-irradiated region. Raman spectra obtained from regions irradiated at lower fluences are presented for (d) 697 J/cm$^2$, (e) 273 J/cm$^2$, (f) 164 J/cm$^2$, (g) 95 J/cm$^2$, and (h) 55 J/cm$^2$. (i, j) Dependence of peak positions on laser fluence. FWHM dependence on fluence for peaks attributed to (k) $Sb_mTe_3$ ($m$ = 1, 2) subsystem, (l) $GeTe_{4-n}Ge_n$ ($n$ = 0–3) tetrahedra and Te-Te- or octahedral Ge-related vibrations defect modes. The yellow highlights mark the characteristics of the vibrational modes of the initial (non-irradiated) spectra.

Additional experiments of the structural analysis of the post-irradiated areas after exposure to a high-intensity laser source using a single long pulse (0.5 ms) are presented in Figure 7. The data clearly show that the applied irradiation induces localized molten regions on the film surfaces, indicating a thermally dominated modification regime. In contrast, no rod-like structures, observed after ultrafast laser processing (Figure 5), are formed in the $Ge_3Sb_2Te_6$ film, highlighting the difference in the underlying interaction mechanisms. Elemental analysis reveals a slight decrease in the atomic concentration of germanium, which may be attributed to preferential redistribution, diffusion, or possible evaporation under high local temperatures. Elemental distribution profiles are shown in Figure S2(c). XRD analysis indicates that the overall crystalline phase remains unchanged; however, a noticeable shift in lattice parameters is observed, likely associated with compositional variations such as germanium depletion. Raman spectroscopy further supports these findings. For the $Ge_3Sb_2Te_6$ film, the Raman spectrum remains largely similar to that of the as-deposited material, except for an additional peak attributed to octahedrally coordinated Ge. Moreover, the increased visibility of the $A^2_{1g}$ modes $Sb_mTe_3$ ($m$ = 1, 2) subsystem suggests the formation of structural defects and partial suppression of vibrational contributions associated with the GeTe sublattice. In comparison, the Raman spectra of $Ge_2Sb_2Te_5$ and $GeSb_2Te_4$ films exhibit a more pronounced suppression of $GeTe_{4-n}Ge_n$ ($n$ = 0–3) tetrahedra vibrational modes, consistent with the trends observed in the compositional and structural analyses.

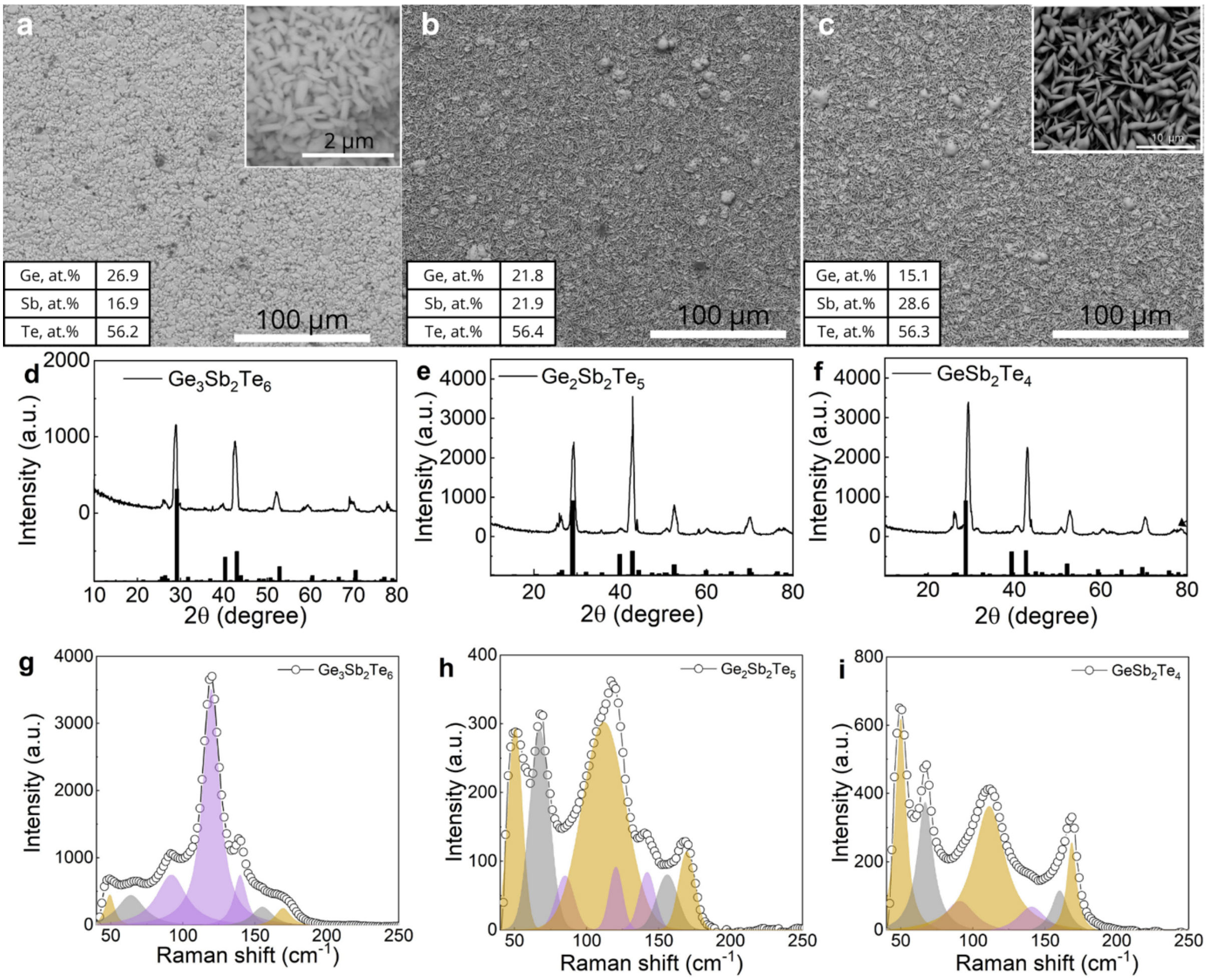

Figure 7. Structural analysis of post-irradiated areas. SEM images of (a) $Ge_3Sb_2Te_6$, (b) $Ge_2Sb_2Te_5$, and (c) $GeSb_2Te_4$, including corresponding elemental atomic distributions. Insets show higher-magnification views of the surface morphology. XRD analysis with corresponding reference PDF-2 database cards for (d) $Ge_3Sb_2Te_6$, (e) $Ge_2Sb_2Te_5$, and (f) $GeSb_2Te_4$. Raman spectra with fitted Lorentz peaks for (g) $Ge_3Sb_2Te_6$, (h) $Ge_2Sb_2Te_5$, and (i) $GeSb_2Te_4$. Peak assignments are indicated by color: purple- $GeTe_{4-n}Ge_n$ ($n$ = 0–3) subsystem, yellow- $Sb_mTe_3$ ($m$ = 1, 2) subsystem, and gray-Te–Te or defective octahedral Ge vibrational modes.

The different optical responses of $GeSb_2Te_4$, $Ge_2Sb_2Te_5$, and $Ge_3Sb_2Te_6$ after laser irradiation demonstrate that the GST compositions provide distinct pathways for spectral tuning, which is attributed to the interplay of their compositional, structural and morphological characteristics. Both $Ge_2Sb_2Te_5$ and $GeSb_2Te_4$ exhibit absorbance enhancement after irradiation, and both possess anisotropic plate-like crystallites oriented perpendicular to the substrate, suggesting that this morphology plays a role in the enhanced optical response, either through light scattering effects or through laser-induced modification of the crystallite interfaces. In particular, $GeSb_2Te_4$, which has the largest plates, the narrowest band gap and the smallest film thickness, exhibits the strongest modulation in the NIR range. In contrast, $Ge_2Sb_2Te_5$, with its well-defined plate geometry and intermediate band gap, shows the highest optical contrast in the Vis range, indicating a stronger sensitivity of this composition to laser-induced modifications of the local bonding environment. Meanwhile, $Ge_3Sb_2Te_6$ exhibits a qualitatively different laser response, characterized by a decrease in absorbance across both the Vis and NIR spectral ranges, which is probably consistent with the coupled effect of laser-induced structural reorganization of the GeTe sublattice, the major structural contributor in this composition, the equiaxed fine-grained morphology, and the concurrent surface morphological evolution at high fluences, where the fine grains transform into rod-like structures.

Overall, these results highlight that the laser response of GST films is composition-dependent and that controlled selection of the GST composition provides a route for selective modification of optical properties in different spectral ranges. No phase transition occurs; instead, the material undergoes a controlled structure-driven reconfiguration governed by the irradiation conditions.

**Conclusion**

This work establishes a direct link between composition, microstructure, and the broadband 400-1700 nm optical response across a compositional gradient deposited in a single CVD process. The crystalline GST films investigated here exhibit a composition-dependent optical response that can be selectively tuned by femtosecond laser irradiation, a capability not accessible through conventional amorphous–crystalline phase-change approaches. The results showed that optical behavior is critically governed by the films' composition and microstructure of their constituent crystallites. Along the deposition gradient from $GeSb_2Te_4$ to $Ge_3Sb_2Te_6$, crystallite morphology evolved from anisotropic plate-like crystallites with aspect ratios of 9.0 and 4.7 for $GeSb_2Te_4$ and $Ge_2Sb_2Te_5$, respectively, to grains of 0.2 μm in $Ge_3Sb_2Te_6$. This structural transition was accompanied by an increase in the GeTe structural units, confirmed by Raman spectroscopy, where $GeTe_{4-n}Ge_n$ ($n$ = 0–3) tetrahedral modes showed the highest sensitivity to compositional variations. At the same time, the optical absorbance minimum shifted from 815.9 to 889.8 nm along the deposition gradient, accompanied by a 5.9-fold change in absorbance intensity within the 400-1000 nm spectral range. Among the individual films, $Ge_3Sb_2Te_6$ exhibited the highest spatial variations in spectral shifts along both *L*- and *T*-directions compared with $GeSb_2Te_4$ and $Ge_2Sb_2Te_5$. This variation reflects contributions from both the compositional gradient and morphology-related effects, though their individual contributions could not be quantitatively separated in the present study. Local femtosecond laser irradiation produced a fluence-dependent increase in absorbance for $Ge_2Sb_2Te_5$ and $GeSb_2Te_4$. The highest NIR optical contrast (3.2-fold) was achieved for $GeSb_2Te_4$ at a fluence of 191 J/cm$^2$, whereas the largest visible contrast (8.1-fold) was obtained for $Ge_2Sb_2Te_5$ at 179 J/cm$^2$. In contrast, $Ge_3Sb_2Te_6$ exhibited a decrease in absorbance after laser irradiation. Raman analysis of the irradiated $Ge_3Sb_2Te_6$ film revealed progressive broadening of *E*-mode peaks associated with both $Sb_mTe_3$ ($m$=1, 2) and $GeTe_{4-n}Ge_n$ ($n$ = 0–3) subsystems, accompanied by shifts of Te-Te- or octahedral Ge-related modes. At fluences of 697 J/cm$^2$ and above, these structural changes were accompanied by the rod-like structures formation. The laser response of GST films is strongly composition-dependent, enabling selective modification of optical properties in Vis-NIR spectral ranges. This composition-selective tunability extends the applicability of GST films beyond phase-change memory toward adaptive optical coatings, selective absorbers, and spatially programmable photonic elements.

# Supplementary Information

## Laser-Induced Optical Reconfiguration in Ge–Sb–Te Films with Composition-Dependent Response

M. Zhezhu[1,†], A. Vasil'ev[1], M. Sukiasyan[2], A. Kutuzyan[2], N. Anosov[3], M. Yaprintsev[3], D. A. Ghazaryan[4], H. Gharagulyan[1,2†]

[1] *Liquid Crystalline Nanosystems Research Group, A.B. Nalbandyan Institute of Chemical Physics NAS RA, Yerevan 0014, Armenia*

[2] *Institute of Physics, Yerevan State University, Yerevan 0025, Armenia*

[3] *Belgorod State University, Belgorod 308000, Russia*

[4] *Laboratory of Advanced Functional Materials, Yerevan State University, Yerevan 0025, Armenia*

[†] *The correspondence should be addressed to:*

*marina.zhezhu@ichph.sci.am and herminegharagulyan@ysu.am*

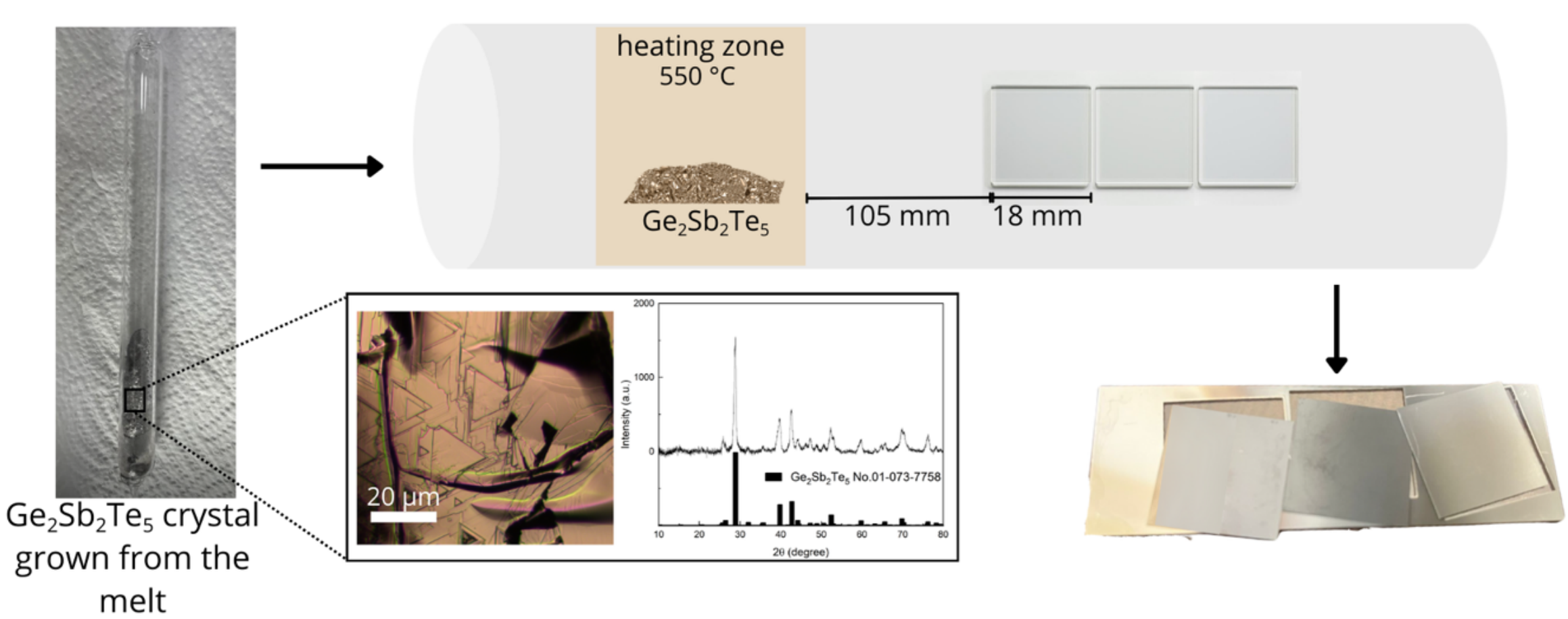

Figure S1. Schematic illustration of the film deposition procedure, involving the image of the as-grown $Ge_2Sb_2Te_5$ crystal used as a source material for film deposition *via* chemical vapor deposition

Table S1 Femtosecond laser irradiation parameters applied to GST films: number of pulses, laser power and pulse energy

| System | Repetition rate, MHz | Irradiation time, s | Number of pulses | Laser power, W | Pulse energy, J |
|---|---|---|---|---|---|
| $Ge_3Sb_2Te_6$ | 76 | 2 | $1.5 \cdot 10^8$ | 0.137 | $1.8 \cdot 10^{-9}$ |
| | | 3.5 | $2.7 \cdot 10^8$ | | |
| | | 6 | $4.6 \cdot 10^8$ | | |
| | | 10 | $7.6 \cdot 10^8$ | | |
| | | 5 | $3.8 \cdot 10^{8}$ | 0.7 | $9.2 \cdot 10^{-9}$ |
| | | 30 | $2.3 \cdot 10^{9}$ | | |
| $Ge_2Sb_2Te_5$ | 80 | 0.8 | $6.4 \cdot 10^{7}$ | 0.6 | $7.5 \cdot 10^{-9}$ |
| | | 1 | $8 \cdot 10^{7}$ | | |
| | | 1.5 | $1.2 \cdot 10^{7}$ | | |
| | | 1.9 | $1.5 \cdot 10^{7}$ | | |
| $GeSb_2Te_4$ | 76 | 1 | $7.6 \cdot 10^{7}$ | *0.137* | $1.8 \cdot 10^{-9}$ |
| | | 3 | $2.3 \cdot 10^{8}$ | | |
| | | 4.5 | $3.4 \cdot 10^{8}$ | | |
| | | 7 | $5.3 \cdot 10^{8}$ | | |

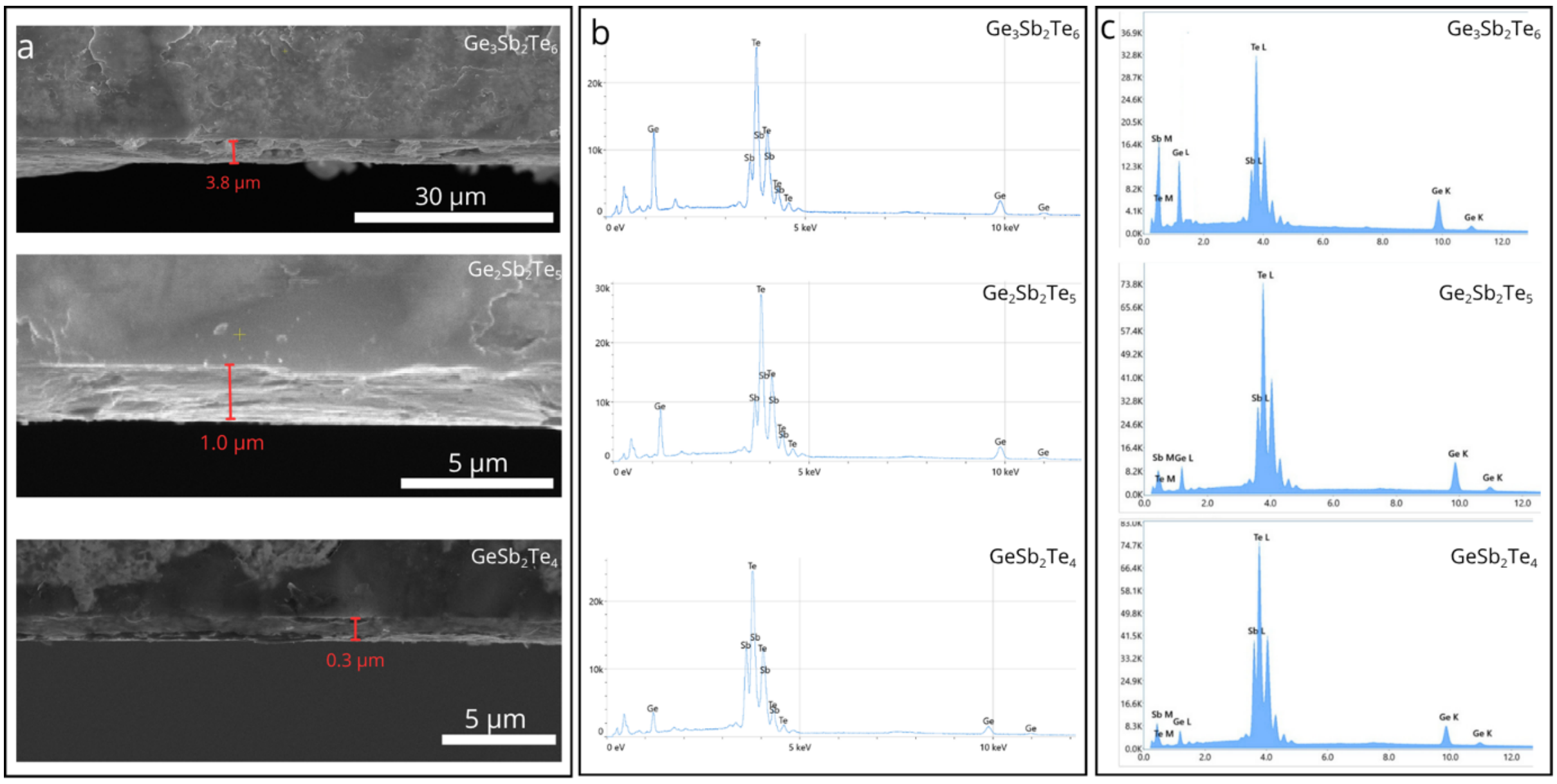

Figure S2. (a) Cross-sectional SEM images of Ge-Sb-Te films with corresponding film thicknesses. The red values indicate the average thickness obtained from a log-normal distribution analysis of at least 30 measurements collected from several images. (b) Elemental distribution profiles of the as-deposited Ge-Sb-Te films. (c) Elemental distribution profiles of the laser-irradiated regions of the Ge-Sb-Te films.

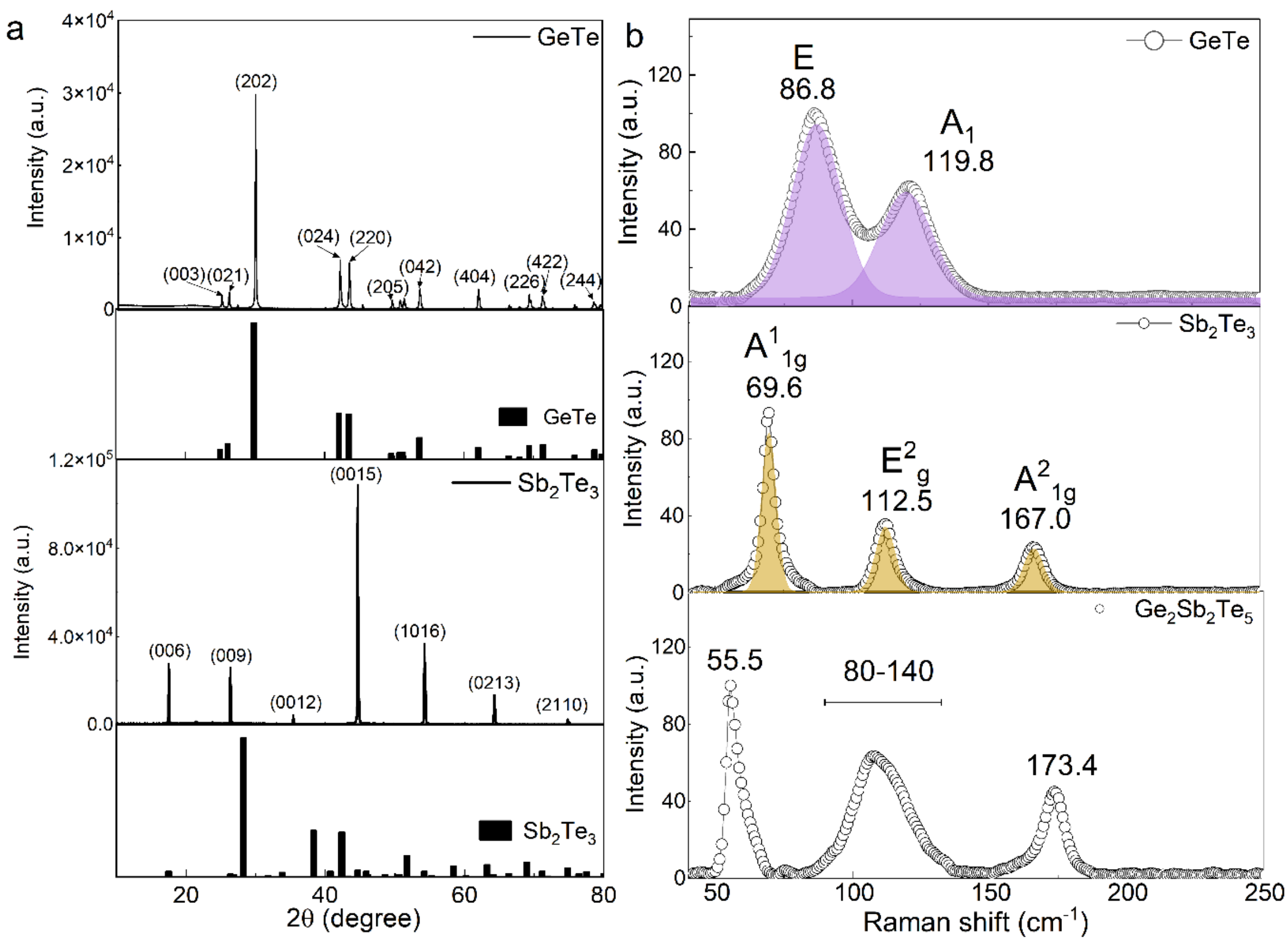

Figure S3. Structural analysis of GeTe and $Sb_2Te_3$ crystals synthesized by solid-state reaction from high-purity elements in sealed quartz ampoules: (a) XRD patterns with corresponding PDF-2 database cards: Card No. 01-077-3540 for GeTe and Card No. 01-086-7466 for $Sb_2Te_3$. The XRD patterns are accompanied by indexed crystallographic planes, demonstrating the preferred orientation of the $Sb_2Te_3$ crystal along the (00*c*) planes. (b) Room-temperature Raman spectra of the GeTe and $Sb_2Te_3$ crystals, as well as the $Ge_2Sb_2Te_5$ crystal used as a source material for film deposition.

*Detailed Characterization of the Optical Response of GST films.*

*As-deposited GST films*

$Ge_3Sb_2Te_6$ exhibits a decrease in absorption from $A \approx 1$ at 500 nm to an absorption minimum at 884.5 nm ($A \approx 0.39$). After that, the absorbance remains around 0.4, and a second minimum is observed at 1656.1 nm with an absorption of 0.39. The $Ge_2Sb_2Te_5$ film exhibits an absorption minimum at 823.1 nm ($A \approx 0.1$). This is followed by an increase up to 1033 nm ($A \approx 0.41$), and then a broad spectral feature with a peak at 1370.8 nm and an absorbance of 0.39. Within the 1000–1700 nm region, a non-monotonic spectral response is observed. The $GeSb_2Te_4$ displays an absorption minimum at 828.8 nm ($A \approx 0.1$). After 1011 nm, a less pronounced peak is observed at approximately 1381.5 nm with an absorbance of 0.3.

*Laser-irradiated GST films*

The optical absorption spectra of the $GeSb_2Te_4$ film no longer exhibit a distinct absorption minimum in the range of 815-840 nm; instead, an absorption maximum appears at approximately 1400 nm. For this film, structural transformations are initiated at a laser fluence of 27 J/cm$^2$, accompanied by pronounced changes in the parameters of the absorption band. With increasing laser fluence, the absorption peak gradually shifts toward longer wavelengths and exhibits progressive broadening. At the highest fluence, a substantial increase in the peak width is observed, indicating significant structural modifications in the material and a pronounced broadening of the absorption band.

The spectra recorded after irradiation of $Ge_2Sb_2Te_5$ film in the Vis range were significantly altered, with no noticeable absorption observed up to approximately 560 nm, followed by strong fluctuations extending to about 740 nm. In the 750–1100 nm region, the films exhibited high absorbance, while the absorption peak near 1600 nm remained preserved, similarly to the initial spectrum. With increasing laser fluence, the peak position remained nearly unchanged, whereas the FWHM gradually increased, indicating progressive broadening of the absorption band. Thus, laser irradiation suppressed $A_{min}$ near 820 nm and enhanced the overall absorbance of the film. At the same time, the optical behavior in the 1000–1700 nm range was largely retained, with the primary effect being an increase in absorbance intensity accompanied by band broadening.

Spectra recorded after $Ge_3Sb_2Te_6$ film irradiation demonstrated a distinct optical behavior. Compared to other compositions, this film exhibits significant changes in the 400–1000 nm region while maintaining similar behavior in the 1000–1700 nm range. Initially characterized by high absorption in the 400–1000 nm region, with a minimum at ∼820 nm, laser irradiation results in the disappearance of this minimum and a subsequent monotonic decrease in absorption up to 1660 nm, where another minimum is observed. After laser irradiation, $\lambda_{min}$ shifted from 1658.6 nm (for as-deposited film) to 1660.6-1659.9 nm, depending on the irradiation conditions. At the same time, the minimum remained nearly unchanged, with the FWHM showing no significant variation. A further increase in fluence, up to 4180 J/cm$^2$, led to the complete suppression of $\lambda_{min}$.

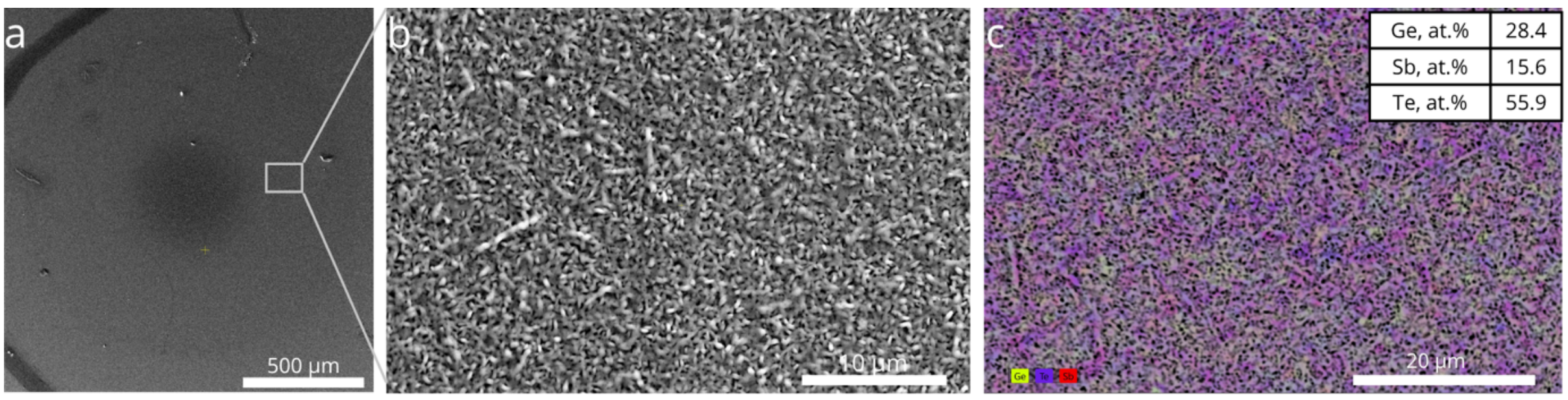

Figure S4. Structural analysis of the laser-irradiated region at 4180 J/cm$^2$. (a) SEM image of the entire irradiated surface, illustrating the induced morphological changes. (b) Magnified view of the edge region. (c) Elemental distribution maps of Ge, Sb, and Te, together with the corresponding atomic concentrations of the elements in the irradiated spot's center.

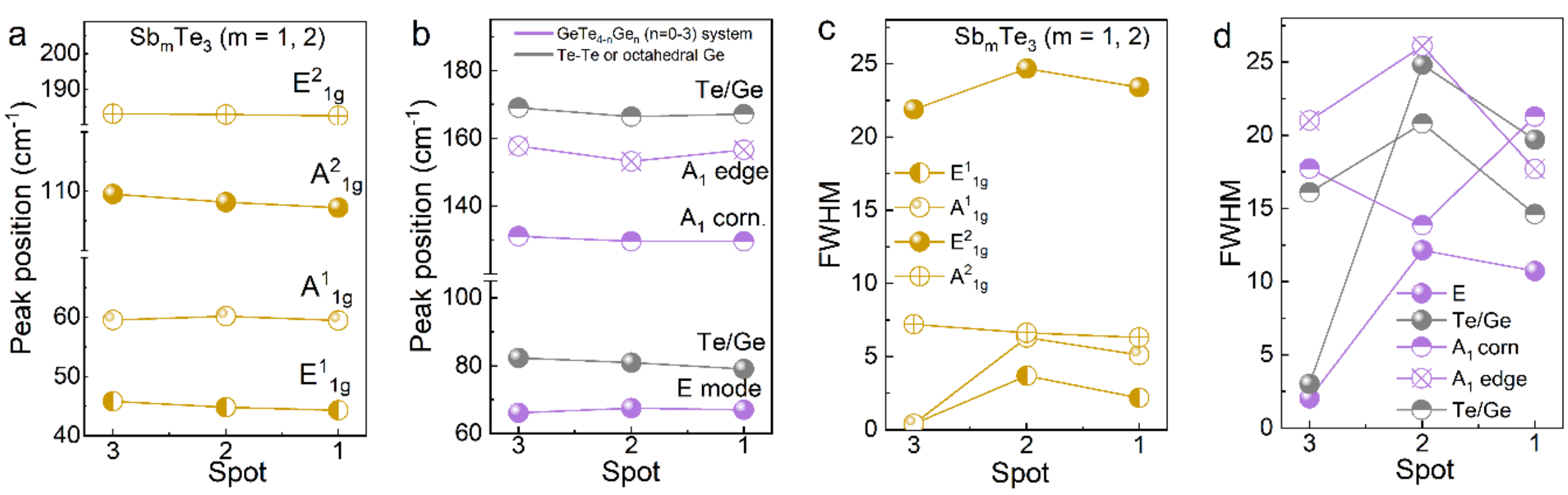

Figure S5. Detailed analysis of peak positions for three representative spots in the 4180 J/cm2 lase fluence case. Along the x-axis, positions 1, 2 and 3 correspond to the labeled spectra and represent the center, edge, and boundary regions, respectively. Dependence of FWHM on laser fluence for peaks attributed to the (c) $Sb_mTe_3$ (*m* = 1, 2) subsystem and (d) $GeTe_{4-n}Ge_n$ (*n* = 0–3) tetrahedra, as well as Te-Te- and octahedral Ge-related defect vibration modes.